\documentclass[letterpaper,journal]{IEEEtran}
\usepackage{amsmath,amsfonts,amssymb}
\usepackage{graphicx}
\usepackage{booktabs}
\usepackage{dblfloatfix}
\usepackage{placeins}

\usepackage{url}
\usepackage[hidelinks]{hyperref}
\newcommand{\reproguide}{Sec.~\ref{sec:reproducibility}}

\title{Confoundry: Spatial Causal Discovery and Dynamic Effect Mapping from Earth Observation Time Series}
\author{Vytautas Jancauskas, Daniela Espinoza-Molina, and Samuel Garske%
\thanks{Corresponding author: V. Jancauskas.}%
\thanks{The authors are with the EO Data Science, Earth Observation Center (EOC), German Aerospace Center (DLR), Germany (e-mail: vytautas.jancauskas@dlr.de).}%
\thanks{V. Jancauskas's ORCID is \href{https://orcid.org/0000-0002-6051-210X}{0000-0002-6051-210X}.}%
\thanks{This work was funded in part under the Sustainable Blue Economy Partnership's
Second Joint Transnational Call 2024, SBEP2024-31 (Digital Twin for Innovation
in Oceanic Visualization and Exploration -- DIVE).}
}

\begin{document}

\maketitle

\begin{abstract}
Earth observation time series provide extensive historical observations of environmental change, but causal relationships inferred from them may vary geographically and depend heavily on local model adequacy. We introduce Confoundry, a framework for spatially resolved causal discovery and dynamic effect estimation from heterogeneous Earth observation time series. Confoundry aligns multi-source raster data, removes deterministic seasonal and long-term variation, fits separate structural VAR--LiNGAM models at individual grid cells, and propagates bootstrap estimation uncertainty into mapped dynamic responses. Model diagnostics are evaluated spatially so that mapped responses can be restricted to locations meeting predefined model checks. We demonstrate the approach using monthly MODIS NDVI and ERA5-Land hydroclimatic variables over the Iberian Peninsula from 2005--2024. Of 27,050 fitted pixel models, 14,657 passed the primary diagnostic criteria. The retained models yield geographically heterogeneous model-implied NDVI responses to precipitation, temperature, potential evaporation, and shallow and deep soil moisture. Aggregated responses vary systematically along an external aridity gradient: water-related responses become relatively stronger toward drier environments, whereas temperature responses increase toward wetter conditions. A single-pixel synthetic benchmark illustrates substantial finite-record variation in recovered causal structure and dynamic effects from 20-year monthly time series. The results demonstrate that spatial causal mapping of Earth observation data is most informative when estimated responses, uncertainty, and local model adequacy are reported together.
\end{abstract}

\begin{IEEEkeywords}
Earth observation, causal discovery, causal effect estimation, drought, remote sensing time series, structural vector autoregression, VAR--LiNGAM.
\end{IEEEkeywords}

\section{Introduction}

Earth observation (EO) data and reanalysis products provide spatially extensive records of vegetation, climate, and land-surface conditions. They support applications including drought monitoring, crop assessment, ecosystem analysis, and land-surface characterization. Many EO studies describe statistical associations or predict environmental outcomes without identifying causal mechanisms. For example, a correlation between vegetation greenness and precipitation cannot by itself distinguish whether the observed covariation is the result of a direct response, shared seasonality, an indirect soil-moisture pathway, or a common driver of both variables.

Causal inference provides a way to formulate these alternatives as directed relationships with explicit assumptions \cite{Pearl2009,imbens2015causal,scholkopf2021toward}. This can help separate direct and mediated effects and generate testable hypotheses about environmental processes. Observational causal estimates remain conditional on the variables measured and the assumptions of the model; they complement domain knowledge and process-based modeling rather than replacing them.

EO data make such analyses difficult in several ways. Products differ in grid, resolution, cadence, data availability, and physical meaning. Relationships also vary across landscapes, so a single regional graph may conceal local mechanisms. Finally, environmental systems can violate assumptions commonly used in causal discovery, including linearity, independent disturbances, stable dynamics, and adequate control of common causes. Spatial causal analysis of EO data therefore requires more than applying a causal-discovery algorithm to each grid cell: it must reconcile heterogeneous observations, represent geographic variation in causal dynamics, and identify where the fitted model itself does not support causal interpretation.

Two relevant approaches to causal research involving EO data are design-based studies and systems-oriented time-series analysis. In design-based studies, satellite products serve as outcomes, exposures, or covariates in cross-sectional experiments and quasi-experiments \cite{jain2020satellite,jerzak2022imageconfounding,jerzak2023imageheterogeneity}. However, EO data do not themselves provide causal identification, and retrieval error, spatial scale, and post-treatment information can bias otherwise valid analyses \cite{jain2020satellite,sakamoto2024scoping,vancleemput2025scaling}.

Conversely, Earth-system research often seeks causal structure among coupled observational time series. Approaches range from Granger causality and convergent cross mapping \cite{granger1969investigating,sugihara2012detecting} to conditional-independence methods such as PCMCI \cite{ebertuphoff2012causal,perezsuay2019geoscience,runge2018causal,runge2019detecting}. These approaches have been applied to climate teleconnections and biosphere--atmosphere interactions \cite{runge2015gateways,nowack2020causal,krich2020estimating}. Methods based only on lagged predictability do not separately identify contemporaneous direction, whereas contemporaneous discovery requires additional assumptions. Furthermore, reviews emphasize that autocorrelation, seasonality, non-stationarity, hidden common causes, and short records make causal interpretation strongly assumption-dependent \cite{runge2019causation,runge2023timeseries}. Spatial confounding and interference add further identification problems that ordinary pixel-wise regression does not resolve \cite{paciorek2010scale,reich2021spatial}.

To examine local heterogeneity in both contemporaneous and delayed dependencies, we introduce Confoundry, a pixel-wise causal-analysis framework. We fit a separate structural vector autoregressive linear non-Gaussian acyclic model (VAR--LiNGAM) at each grid cell \cite{hyvarinen2010svar,shimizu2011directlingam}. Under its structural assumptions, the non-Gaussian error model can identify a contemporaneous causal order that lag-only models cannot recover. Fitting this model locally allows contemporaneous structure, lagged relationships, effect magnitude, and response persistence to vary across the study area. The resulting quantities remain spatially linked to the input raster grid and can therefore be mapped together with model diagnostics and uncertainty. Confoundry integrates established structural estimation methods into a reproducible spatial workflow; it does not introduce a new causal identification algorithm. This local flexibility trades spatial pooling for higher data requirements and does not account for causal spillovers between neighboring pixels.

We demonstrate the framework using drought-related vegetation dynamics across the Iberian Peninsula. Vegetation can respond to precipitation, temperature, evaporative demand, and water stored at different soil depths, with pathways that differ by location. Monthly MODIS NDVI and five ERA5-Land variables from 2005--2024 are adjusted for local calendar-month seasonality and linear trend. We fit structural VAR--LiNGAM to the six residualized series at each eligible pixel, estimate NDVI responses to one-time hydroclimatic disturbances through month 12, and assess estimation uncertainty through paired graph-bootstrap replicates.

The study asks three questions: how hydroclimatic effects on vegetation vary across space and response horizon; where the fitted local models satisfy the evaluated diagnostic criteria; and whether the resulting geographical variation is consistent with an independent climatic aridity gradient. We report dynamic response trajectories, cumulative-effect maps, spatial model diagnostics, synthetic recovery experiments, and a comparison with external aridity climatology. Reproduction commands and input configurations are provided in \reproguide.

The main methodological contributions in this paper are as follows:
\begin{itemize}
    \item We formulate EO causal analysis as a set of local structural time-series models, allowing contemporaneous causal structure, lagged relationships, and dynamic effects to vary geographically rather than imposing one graph over the full study domain.

    \item We translate fitted structural VAR--LiNGAM models into horizon-specific and cumulative response maps, using local variability-based scaling to make responses across heterogeneous variables and locations interpretable on a common relative scale.

    \item We report causal responses together with local diagnostics of structural-error dependence, residual temporal dependence, and dynamic stability, and propagate paired graph-bootstrap uncertainty through the estimated response dynamics.

    \item We assess finite-record recovery using synthetic time series generated under known structural models and evaluate the geographical plausibility of the inferred vegetation sensitivities against an independent aridity climatology.
\end{itemize}

The resulting maps are model-conditional causal hypotheses, not substitutes for experimental identification or process-based simulation. They provide a spatially explicit basis for further evaluation of hydroclimatic pathways.

\section{Case study: drought in the Iberian Peninsula}

The Iberian Peninsula has a diverse range of climates and biomes. Variable precipitation, high evaporative demand, and pressure on water resources make drought impacts on vegetation a significant concern  \cite{almendra_martin_agricultural_2021,gaona_interactions_2022,soares_future_2023}. This diversity also makes the region useful for examining whether vegetation responses to hydroclimatic anomalies actually differ across locations.

\subsection{Data sources}

\begin{table}[!tbp]
\centering
\caption{Variables used in the six-dimensional VAR--LiNGAM model.}
\label{tab:data_inputs}
\small
\renewcommand{\arraystretch}{1.12}
\begin{tabular}{@{}lll@{}}
\toprule
Symbol & Physical quantity & Product field \\
\midrule
$T$ & 2-m air temperature & ERA5-Land \texttt{t2m} \\
$P$ & Total precipitation & ERA5-Land \texttt{tp} \\
$E$ & Potential evaporation & ERA5-Land \texttt{pev} \\
$SM_s$ & Soil moisture (7--28\,cm) & ERA5-Land \texttt{swvl2} \\
$SM_d$ & Soil moisture (28--100\,cm) & ERA5-Land \texttt{swvl3} \\
$N$ & NDVI & MODIS MOD13C2 \\
\bottomrule
\end{tabular}
\end{table}

\begin{figure}[!tbp]
    \centering
    \includegraphics[width=\linewidth]{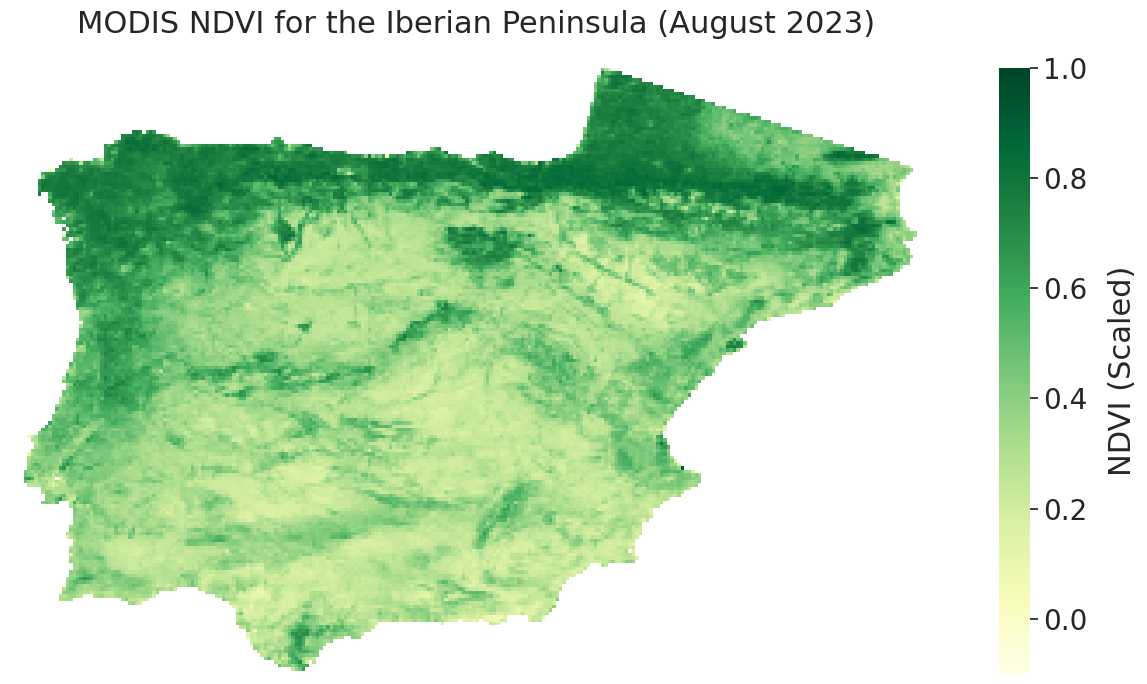}
    \caption{MODIS MOD13C2 NDVI over the Iberian Peninsula in August 2023. Values are shown after application of the product scale factor of $10^{-4}$.}
    \label{fig:iberian_peninsula_ndvi}
\end{figure}

The study region follows the boundary in \texttt{data/iberian\_peninsula.json} of the Confoundry repository (see \reproguide) and is shown in Fig.~\ref{fig:iberian_peninsula_ndvi}. The dataset spans January 2005 to December 2024 at a monthly resolution. Table~\ref{tab:data_inputs} lists the six variables used by the VAR--LiNGAM model. The ERA5-Land data comprise 2-m air temperature, total precipitation, potential evaporation, and volumetric soil water in the 7--28\,cm and 28--100\,cm layers \cite{munozsabater2021era5land}. Vegetation greenness is represented by the MODIS MOD13C2 monthly normalized difference vegetation index (NDVI) on its 0.05$^{\circ}$ reference grid \cite{didan2021mod13c2}. ERA5-Land \texttt{pev} estimates open-water (pan) potential evaporation, not actual evapotranspiration. Greater potential evaporation is recorded as more negative \texttt{pev}. We preserve this sign: a positive \texttt{pev} pulse represents \emph{reduced} potential evaporation \cite{ecmwf_era5_land_docs}. Acquisition and grid alignment procedures are described in Methods and \reproguide.

Each series was adjusted for calendar-month (seasonal) effects and a linear trend before causal modeling. The fitted baseline accounts for regular seasonal variation; seasonality was not entered as a node in the VAR--LiNGAM model. All six residualized variables refer to the same calendar month at time $t$. 

\subsection{Conceptual drought pathways}

\begin{figure*}[!tp]
  \centering
  \includegraphics[width=\linewidth]{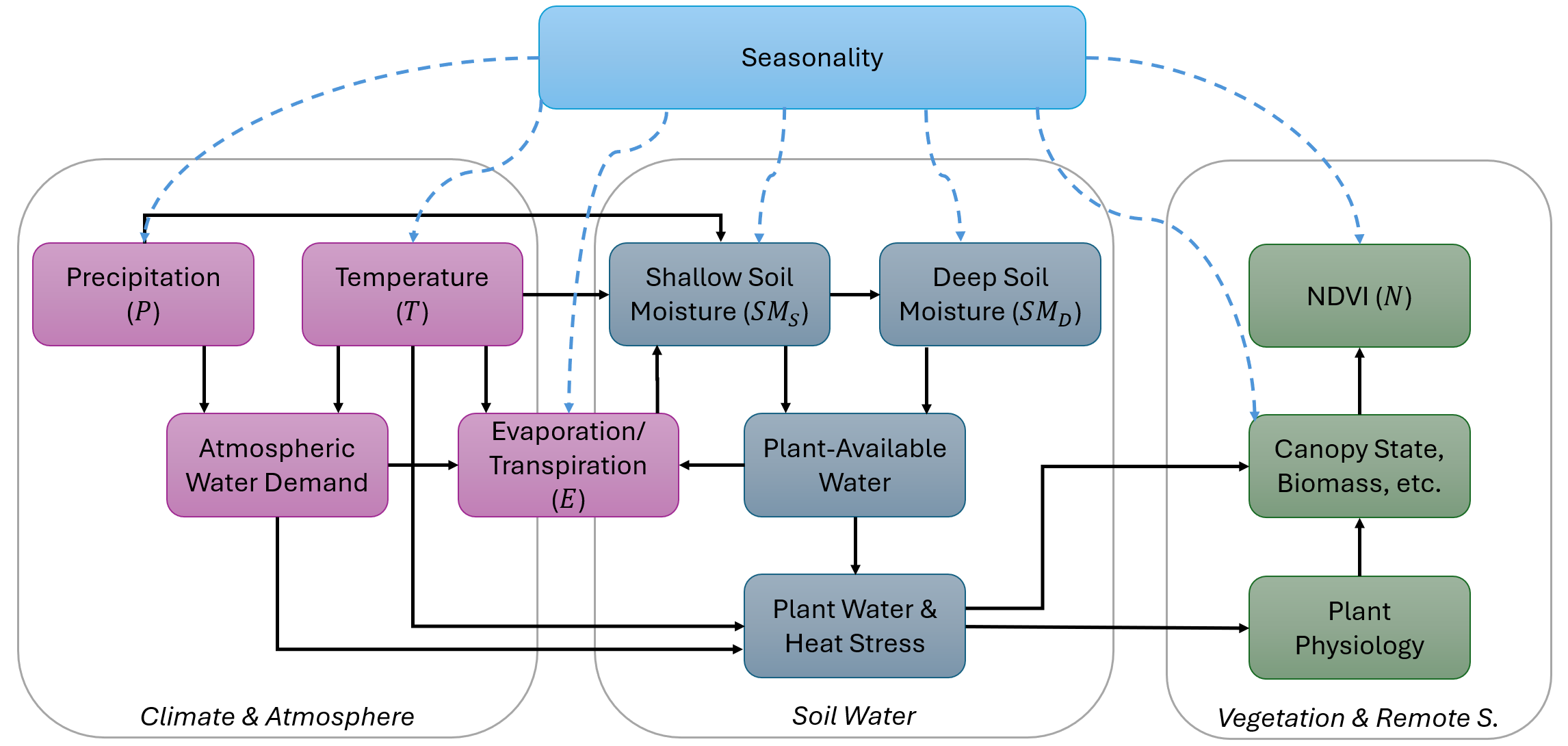}
  \caption{Conceptual ecohydrological pathways linking seasonal forcing, hydroclimatic conditions, plant stress, canopy state, and NDVI. This diagram motivates variable selection; its full structure is not estimated as a graph. Its evaporation/transpiration node denotes an actual flux, not the ERA5-Land potential-evaporation input to the fitted model.}
  \label{fig:dag}
\end{figure*}

Figure~\ref{fig:dag} summarizes plausible ecohydrological pathways motivating the six model variables. Precipitation replenishes soil water, which may redistribute between shallow and deeper layers and affect plant water availability. Temperature and atmospheric demand can affect water stress and canopy development; vegetation can, in turn, alter actual evapotranspiration through canopy conductance and leaf area \cite{allen1998crop,laio2001plants,seneviratne2010soil,grossiord2020plant}. NDVI measures canopy greenness rather than water stress or photosynthesis directly \cite{huete2002overview}.

The diagram includes actual evaporation/transpiration as a physical flux, although the fitted model instead uses the signed ERA5-Land potential-evaporation variable \texttt{pev}. The illustration labels that conceptual flux $E$, as does the model equation for \texttt{pev}; the two quantities should not be equated. It also includes processes, such as physiological stress and land management, that are not explicitly observed. The conceptual feedback pathways may unfold at different times and do not imply that the fitted contemporaneous graph contains cycles. Calendar-month effects and linear trend were removed before fitting; neither was treated as a causal graph node. Thus, the diagram motivates variable selection but does not establish the completeness or correctness of the estimated local graphs \cite{runge2019causation,runge2023timeseries}.

\subsection{Pixel-wise dynamic modeling of drought--vegetation pathways}

Drought effects on vegetation reflect interacting precipitation, temperature, evaporative demand, and soil-water processes. We examine how the fitted relationships between these variables and NDVI vary over the Iberian Peninsula. The model contains six residualized variables observed in the same month:

\begin{equation}
    \mathbf{z}_{it} = [P_{it},T_{it},E_{it},SM_{s,it},SM_{d,it},N_{it}]^{\mathsf T},
\end{equation}
where $i$ indexes pixels; $P$ is precipitation, $T$ temperature, $E$ potential evaporation, $SM_s$ and $SM_d$ soil moisture in the 7--28\,cm and 28--100\,cm layers, and $N$ NDVI. Each component is an anomaly after removal of its local calendar-month pattern and linear trend. The structural VAR--LiNGAM model combines contemporaneous effects with three monthly lags; its estimation and assumptions are specified in Methods.

The guiding question is how hydroclimatic disturbances propagate into vegetation-greenness anomalies, and how the fitted responses differ spatially. We report trajectories and cumulative maps for each source variable. The following interpretation distinguishes the reported impulse responses from hypothetical intervention scenarios, which are not evaluated here.

Comparing cumulative responses across source variables provides a descriptive indication of which fitted hydroclimatic pathways are most influential at a location. Larger absolute precipitation or soil-moisture responses suggest stronger water sensitivity; temperature and potential-evaporation responses may indicate sensitivity to atmospheric conditions. Because the source pulses are scaled locally and effect uncertainty differs between variables, these responses should not be interpreted as a categorical classification of drought regime.

The reported response maps describe model-implied effects of one-time structural disturbances. They do not simulate the consequences of a physical irrigation action, which would also involve water delivery, infiltration, runoff, and management constraints. Intervention scenarios based on the fitted equations are a possible later application, but none is evaluated in the present case study \cite{Pearl2009}.

\subsection{Interpretation and limitations}

The fitted effects have a causal interpretation only under the assumptions of the structural VAR--LiNGAM framework: linearity, an acyclic contemporaneous structure, adequate temporal specification, no relevant hidden confounding, and mutually independent non-Gaussian structural errors \cite{hyvarinen2010svar,shimizu2011directlingam}. In environmental remote sensing, these assumptions are strong. Potential violations include irrigation, land management, soil texture, topography, groundwater access, land-cover heterogeneity, and unobserved human interventions. Fixed site characteristics do not vary within a pixel, but unobserved time-varying drivers and changes in management can still confound the fitted relationships. Per-pixel modeling does not remove these concerns.

The per-pixel graphs and effects should be treated as process-oriented hypotheses rather than definitive proof of causality. The aridity-climatology comparison below assesses external consistency, not identification. Independent measurements, land-cover-stratified checks, or quasi-experimental evidence would be needed to support more specific causal claims.

\section{Methods}

\begin{figure*}[!tp]
    \centering
    \includegraphics[width=\textwidth]{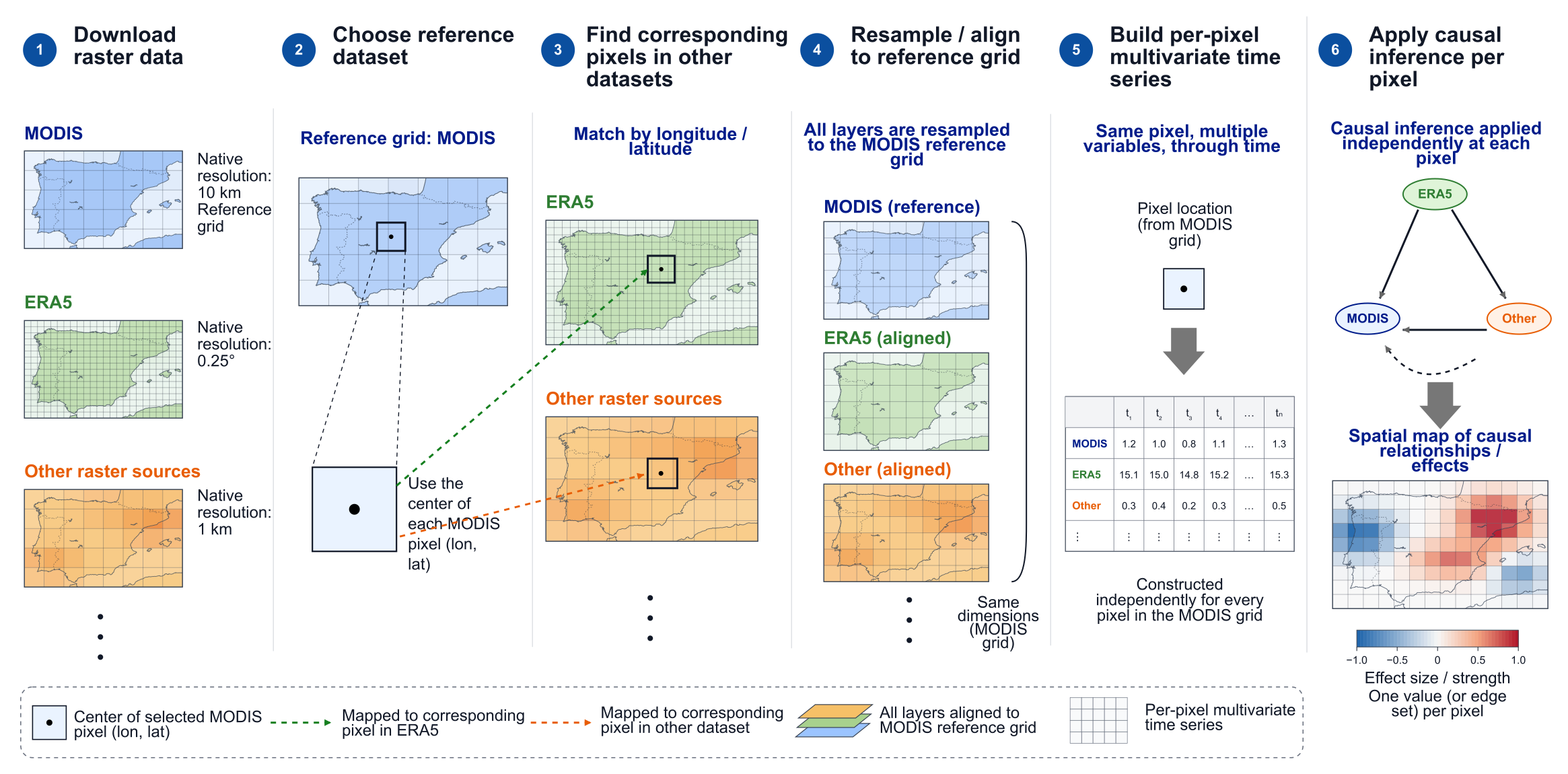}
    \caption{Raster alignment and per-pixel causal inference workflow. The schematic describes grid alignment as resampling; the implemented operation selects each source cell containing a reference-grid cell centre, without interpolation or smoothing.}
    \label{fig:raster-causal-workflow}
\end{figure*}

Figure~\ref{fig:raster-causal-workflow} summarizes the pipeline. The experiment configuration defines the area of interest, source products, date range, reference grid, and processing settings. After acquisition, values from each auxiliary raster are assigned using the source cell containing each MODIS NDVI grid-cell centre. This coordinate-based alignment does not interpolate or smooth source rasters. The aligned monthly fields form a six-variable time series at each reference-grid location.

The remaining stages remove deterministic seasonality and trend, fit an independent structural VAR--LiNGAM model at each eligible pixel, evaluate model diagnostics, and calculate NDVI responses. Section-specific parameters govern the corresponding steps; \reproguide gives the configuration paths and commands used for this study. The common grid preserves a direct spatial link between the observations, fitted models, and output maps; the source products nonetheless retain their native effective spatial resolutions.

\subsection{Study domain and Earth-observation data}

We analyzed monthly environmental variation over the Iberian Peninsula from
January 2005 through December 2024.  The study footprint was defined by the
EPSG:4326 polygon archived with the analysis (see \reproguide).  Five variables were obtained
from the Copernicus Climate Data Store ERA5-Land monthly-mean reanalysis
\cite{munozsabater2021era5land}:
2-m air temperature, total precipitation, potential evaporation, volumetric
soil water in layer 2 (7--28\,cm), and volumetric soil water in layer 3
(28--100\,cm).  Vegetation state was represented by the Terra MODIS MOD13C2
Collection 6.1 monthly normalized difference vegetation index (NDVI), using
the 0.05$^{\circ}$ climate-modeling-grid product distributed by the
Integrated Climate Data Center \cite{didan2021mod13c2}.

Each monthly field was downloaded and clipped to the study polygon.  NetCDF
scale and missing-value metadata were decoded when the source data were
opened, and the resulting fields were stored as GeoTIFFs together with a
DuckDB source catalogue.  The complete source catalogue comprised 240 months
for each of six variables (1,440 variable-month rasters).

\subsection{Spatial and temporal data assembly}

The MODIS NDVI raster defined the analysis grid.  For every month, the centre
coordinate of each NDVI grid cell was transformed to the coordinate reference
system of each covariate raster, when required, and assigned the value of the
containing covariate cell.  Thus, the procedure performed coordinate-based
cell matching rather than interpolation or spatial smoothing.  Values outside
a source raster's extent and source nodata values were represented as missing;
GeoTIFF scale and offset metadata were applied before analysis.  The assembled
long-format table retained year, month, reference-grid row and column, cell
centre coordinates, and the six environmental measurements.  It contained
240 monthly observations at each of 27,086 reference-grid locations, for
6,500,640 rows before complete-case filtering.

\subsection{Removal of deterministic seasonality and trend}
\label{sec:residualization}

Deterministic seasonality and long-term trend can create common temporal
structure that is unrelated to the short-run causal relations of interest.
We therefore residualized each environmental variable separately for every
grid cell. For variable $v$ at grid cell $i$ and month $t$, we fitted the following equation:
\begin{equation}
  x_{ivt}
  = \alpha_{iv}
  + \sum_{m=2}^{12}\gamma_{ivm}
      \mathbb{1}\!\left(M_t=m\right)
  + \beta_{iv}\frac{q_t-\bar q_{iv}}{12}
  + r_{ivt},
  \label{eq:pixel-residualization}
\end{equation}

where $M_t$ is calendar month, January is the reference category, $q_t$ is a
consecutive monthly time index, and $\bar q_{iv}$ is its mean over observations
used to fit that pixel-variable model.  The coefficient $\beta_{iv}$ is
the linear trend per year.  Calendar-month fixed effects were used
instead of a single annual sine--cosine pair so that each variable could have
an arbitrary repeating seasonal profile.  Equation
\ref{eq:pixel-residualization} was estimated by ordinary least squares over the
full 2005--2024 period.  A fit required at least 60 non-missing observations,
at least three observations in every calendar month, and a full-rank design
matrix.  The six residual series $r_{ivt}$, in their original measurement
units, formed the input to causal discovery.  The fitted seasonal and trend
components were not included as endogenous variables.

\subsection{Per-pixel structural VAR--LiNGAM model}

Let $\mathbf z_{it}$ denote the six residualized variables at grid cell $i$ and
month $t$.  A separate structural VAR--LiNGAM model
\cite{hyvarinen2010svar} was fitted at each grid cell:

\begin{equation}
  \mathbf z_{it}
  = \mathbf B_{i0}\mathbf z_{it}
  + \sum_{\ell=1}^{p}\mathbf B_{i\ell}\mathbf z_{i,t-\ell}
  + \mathbf e_{it}.
  \label{eq:structural-varlingam}
\end{equation}

Matrix entries follow the child-by-parent convention, so
$B_{i\ell,jk}$ represents the effect of variable $k$ on variable $j$ at lag
$\ell$.  VAR--LiNGAM assumes linear relations, an acyclic contemporaneous
matrix $\mathbf B_{i0}$, mutually independent non-Gaussian structural errors,
and no unmeasured common causes sufficient to invalidate the recovered causal
order.  Dynamic interpretation additionally requires a stable VAR over the
analyzed period.

The estimator first fitted an intercept-free reduced-form VAR to the residualized observations,

\begin{equation}
  \mathbf z_{it}
  = \sum_{\ell=1}^{p}\mathbf M_{i\ell}\mathbf z_{i,t-\ell}
  + \mathbf u_{it},
  \label{eq:reduced-var}
\end{equation}

and applied DirectLiNGAM \cite{shimizu2011directlingam} to the reduced-form
innovations $\mathbf u_{it}$ to estimate the contemporaneous causal order and
$\mathbf B_{i0}$.  The lagged structural matrices satisfy
$\mathbf B_{i\ell}=(\mathbf I-\mathbf B_{i0})\mathbf M_{i\ell}$.  Adaptive
lasso pruning was then applied jointly to contemporaneous predecessors and all
lagged predictors.  Models were fitted only where complete observations of all
six variables formed a unique, uninterrupted monthly sequence and provided at
least 50 effective observations after lagging.  Of 27,086 assembled grid
cells, 27,050 met these requirements; 36 were excluded because of insufficient
complete data or a break in the monthly sequence.

We used one common lag order across the spatial domain so that every mapped
coefficient and response described the same temporal model.  The production
order was selected before effect estimation using information-criterion and
innovation-diagnostic preflights.  BIC and HQIC favoured $p=1$ in a 132-pixel
audit, while AIC placed $p=1$ and $p=2$ nearly level (the median AIC difference
for $p=2$ relative to $p=1$ was $-0.0017$).  However, the BIC-selected model,
which used $p=1$ at 27,048 of 27,050 fitted cells, rejected analytical
innovation whiteness at 75.98\% of cells.  With otherwise identical settings,
the rejection rate fell to 61.77\% for fixed $p=2$ and 53.58\% for fixed
$p=3$, but increased to 55.67\% for fixed $p=4$.  All candidate specifications
were dynamically stable at every fitted cell, while the median maximum
same-month structural-error correlation increased from 0.130 for $p=3$ to
0.135 for $p=4$.  We therefore fixed $p=3$ as the most parsimonious order at
which the observed improvement in innovation whiteness ceased.  This choice
was based on model diagnostics, not on the estimated NDVI effects.

\subsection{Graph bootstrap and edge summaries}

We generated 500 VAR--LiNGAM bootstrap replicates per grid cell.  The fitted
reduced-form innovation vectors were resampled jointly across variables,
propagated through the fitted VAR to generate a bootstrap time series, and the
complete VAR--LiNGAM and pruning procedure was refitted.  Contemporaneous and
lagged matrices from a replicate were retained as a pair so that uncertainty
propagation preserved their joint fitted structure.

For descriptive graph summaries, an edge was assigned its bootstrap selection
frequency among coefficients with absolute magnitude at least 0.01.  A
consensus graph retained edges with selection frequency at least 0.70 and
absolute point coefficient at least 0.01.  These thresholds affected the
reported consensus topology only.  Dynamic point effects were computed from
the raw adaptively pruned production matrices, and their uncertainty was
computed directly from the paired bootstrap matrices.

\subsection{Model diagnostics and finite-sample whiteness calibration}

We checked how well each pixel model captured the monthly variation in
the six environmental variables. These checks examined the variation
left unexplained by the model: whether errors were correlated between
variables, whether they retained patterns over time, and whether their
distributions met the assumptions of VAR--LiNGAM. We also checked that
the model's response to a temporary disturbance eventually decayed.

We used structural errors from the final pruned causal model to assess same-month
relationships between variables. We summarized these relationships
using the largest absolute Pearson correlation between error series.
Strong correlations would indicate that the model had not fully
separated the sources of variation affecting different variables.

VAR--LiNGAM also relies on non-Gaussian errors. For each variable, we
therefore measured the asymmetry and tail behavior of the error
distribution using skewness and excess kurtosis, and tested normality
with the Jarque--Bera test at $\alpha=0.05$ \cite{jarque1987test}. We reported the fraction
of the six error series that rejected normality, but did not use this
diagnostic to exclude pixels.

We assessed temporal dependence separately because pruning can change
the errors of the causal model. For this check, we refitted a standard
VAR(3), without an intercept, to each complete pixel series. We
examined whether its residuals were correlated with those from earlier
months, both within each variable and between variables. We calculated
one-month autocorrelations and all correlations at lags of up to
12 months. Remaining correlations would indicate temporal structure
that the three-lag model had not captured.

We also used an adjusted multivariate portmanteau test to assess these lagged
correlations jointly across all six variables \cite{hosking1980multivariate,li1981distribution}. To calibrate the test
for the available monthly record, we used a residual bootstrap as the
primary test. At each pixel, we centred the residuals and resampled
complete six-variable residual vectors. This preserved relationships
between variables within a month, and their non-Gaussian distributions,
while removing dependence between months.

We used the resampled residuals to simulate data from the refitted
stable VAR(3), discarded a 200-month burn-in, and refitted VAR(3) to
each simulated series. We then compared the observed test statistic
with those from the simulations. We requested 499 simulations per
pixel and calculated the bootstrap p-value from the valid simulations.
Random seeds were determined by pixel coordinates to make the results
reproducible across parallel runs.

We classified a model as dynamically stable when temporary
disturbances eventually decayed. Mathematically, this required the
spectral radius of its reduced-form companion matrix to be below one.

A pixel was included in the primary effect analysis only when:

\begin{itemize}
  \item the fitted causal model was dynamically stable;
  \item at least 95\% of its paired graph-bootstrap replicates were stable;
  \item all variables had standard deviations greater than $10^{-12}$;
  \item all absolute same-month correlations between structural errors
        were below 0.20;
  \item all absolute one-month residual autocorrelations were below 0.30;
  \item all absolute residual correlations at lags of 1--12 months,
        within and between variables, were below 0.30; and
  \item the bootstrap test was valid and did not reject the absence
        of temporal correlation ($p_{\mathrm{boot}}\geq0.05$).
\end{itemize}

These post-fitting checks did not change the estimated model
coefficients. Temporal-dependence diagnostics specifically assess a separately refitted, unpruned reduced-form VAR, whereas same-month error dependence is computed from the pruned structural fit; thus, the two tests do not evaluate an identical set of residuals. They retained 14,657 pixels (54.18\%) for the primary
analysis. A stricter sensitivity analysis also required the analytical
portmanteau test to give $p\geq0.05$, retaining 12,140 pixels. A broader
sensitivity analysis included all pixels with dynamically stable
fitted models.

\subsection{Dynamic causal effects on NDVI}

We estimated the response of the NDVI anomaly to a pulse in each of the other
five residualized variables over horizons $h=0,\ldots,12$ months.  For each
pixel, let

\begin{equation}
  \mathbf C=(\mathbf I-\mathbf B_0)^{-1},
  \qquad
  \mathbf A_{\ell}=\mathbf C\mathbf B_{\ell}.
\end{equation}

The dynamic response matrices were calculated recursively as

\begin{equation}
  \mathbf G_0=\mathbf C,
  \qquad
  \mathbf G_h
  =\sum_{\ell=1}^{p}\mathbf A_{\ell}\mathbf G_{h-\ell},
  \quad \mathbf G_j=\mathbf 0\ \text{for }j<0.
  \label{eq:dynamic-response}
\end{equation}

For source $X$ and target NDVI $Y$, $G_h[Y,X]$ is the model-implied NDVI
anomaly response $h$ months after a one-unit structural disturbance in the source
equation, holding other structural disturbances at zero and allowing contemporaneous and delayed propagation through the fitted system.  The cumulative effect through horizon $h$ was
$\sum_{s=0}^{h}G_s[Y,X]$.

To compare variables measured on different scales, each pixel-level effect was
also expressed as the response to a local 10th-to-90th-percentile contrast in the residualized source
variable relative to the corresponding residualized NDVI contrast. The positive contrast follows the stored variable sign. For \texttt{pev}, it represents \emph{reduced} potential evaporation; in this linear model an equal increase in potential evaporation produces the opposite NDVI response:

\begin{equation}
  \widetilde G_h[Y,X]
  =G_h[Y,X]
   \frac{Q_{0.9}(X)-Q_{0.1}(X)}
        {Q_{0.9}(Y)-Q_{0.1}(Y)}.
  \label{eq:scaled-effect}
\end{equation}

Point estimates used the raw pruned matrices.  The same dynamic recursion was
applied to every paired graph-bootstrap replicate; singular, non-finite, or
dynamically unstable replicates were excluded.  Pixel-level 95\% intervals
were the 2.5th and 97.5th percentiles of the remaining bootstrap effects.
These intervals quantify graph-estimation uncertainty conditional on the
residualized data and assumed model; they do not include residualization
uncertainty, measurement error, omitted-variable uncertainty, or dependence
between neighboring grid cells.

Across pixels, response trajectories were summarized by the spatial median
and 5th--95th percentile range.  The primary map showed the scaled cumulative
effect through month 12 for each of the five sources, using a common symmetric
color scale whose limit was the 98th percentile of absolute plotted effects.
This robust clipping affected display only.  Stippling identified cells whose
paired-bootstrap 95\% interval excluded zero.  It was interpreted as
conditional bootstrap support and not as a spatially multiplicity-corrected
hypothesis test.

\subsection{Software and reproducibility}
\label{sec:reproducibility}

The analysis was implemented in Python 3.12 using \texttt{lingam} 1.12.2,
DuckDB, NumPy, pandas, Rasterio, xarray, and rioxarray.  The dependency versions are recorded in the repository \texttt{uv.lock}. The reproduction guide documents data acquisition, analysis commands, configurations, and paths to generated databases and figure tables:
\begin{quote}\footnotesize
\url{https://github.com/DLR-MF-DAS/confoundry/blob/main/docs/varlingam_reproduction.md}
\end{quote}
The guide targets the tagged release \texttt{v0.0.1}. Source data may be revised by upstream providers, and rerunning graph bootstraps can produce slightly different Monte Carlo summaries; exact numerical reproduction therefore additionally depends on retaining the original source data and analysis outputs.

\section{Results}

\subsection{Residualized input data}

\begin{figure*}[!tp]
  \centering
  \includegraphics[width=\textwidth]
    {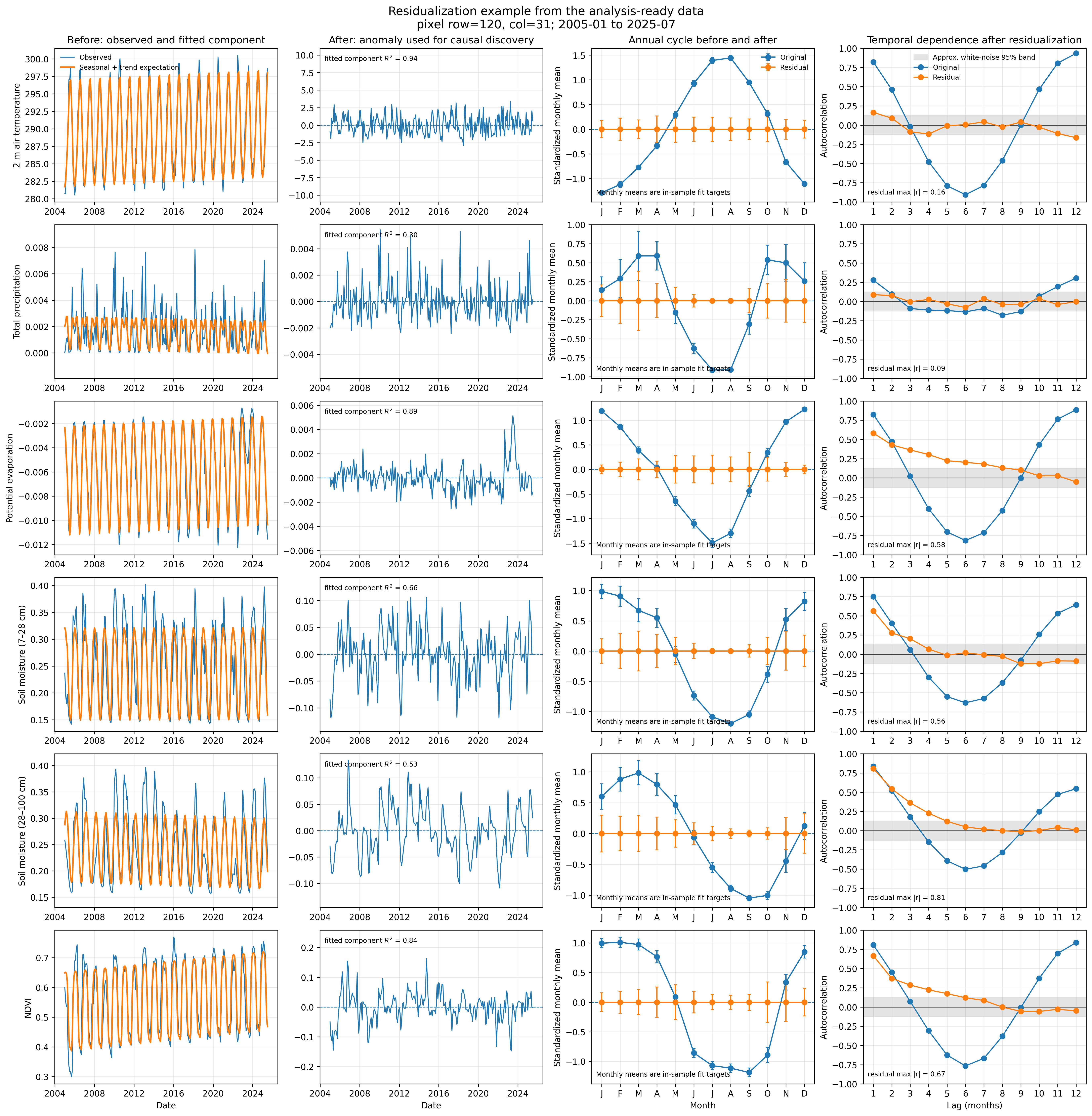}
  \caption{Original and residualized series at a pixel selected for representative residualization performance. Columns show observed values and fitted baseline, residual anomalies, monthly climatology before and after adjustment, and autocorrelation through lag 12.}
  \label{fig:residualized-data}
\end{figure*}

Figure~\ref{fig:residualized-data} illustrates the preprocessing at one
automatically selected pixel. We considered pixels with at least
60 months of complete original values, fitted baselines, and residuals
across all six variables. For each variable at each eligible pixel,
we calculated the fraction of variance removed by preprocessing as
one minus the ratio of residual variance to original variance.
We then averaged these fractions across the six variables and selected
the pixel whose average was closest to the median across eligible
pixels. Ties were resolved by choosing the pixel with more complete
observations, followed by grid row and column.

This selection provides an example with a typical overall reduction
in variance. It does not imply that the pixel represents the region's
environmental conditions or its remaining temporal dependence.
The figure illustrates the transformation at this location rather
than a peninsula-wide pattern. Each location has up to 240 monthly
observations per variable from January 2005 through December 2024.

Seasonal cycles and long-term trends can produce shared variation without the short-term relationships of interest. Such structure can generate spurious time-series associations \cite{granger1974spurious,runge2019causation}. As described in Section~\ref{sec:residualization}, we fit calendar-month effects and a linear trend separately for each variable and pixel. Subtracting this baseline produces the residual anomaly used by VAR--LiNGAM. A positive precipitation residual, for example, means more precipitation than expected at that pixel for the month and date; it does not mean that the location is wetter than another location.

The first column of Figure~\ref{fig:residualized-data} shows each original series with its fitted baseline; the second shows their difference in the original units. The third compares calendar-month averages before and after adjustment. These series are standardized separately, so the vertical axis is in standard deviations, and error bars give the standard error of each monthly mean across years. Since the baseline was fitted to the same observations, the reduced residual seasonality is an in-sample check rather than independent validation.

The rightmost column compares autocorrelations at lags 1--12. The shaded range is an approximate visual reference for uncorrelated observations, not a joint whiteness test; the displayed maximum $|r|$ is the largest absolute residual autocorrelation across those lags. At the illustrative pixel, fitted baselines captured the main repeating cycles and trends, and residual autocorrelations generally moved closer to zero. Some dependence remained. The subsequent VAR represents lagged relationships in the anomalies, while separate innovation diagnostics assess dependence left after model fitting.

Residualization reduces one source of shared temporal variation but cannot establish the causal assumptions of VAR--LiNGAM. In particular, it does not eliminate unobserved time-varying drivers.

\subsection{Spatial variation in model adequacy}
\label{sec:varlingam-qc-results}

We screened each pixel model for stability, sufficient variation in the
input data, and unexplained relationships between variables or across
months (Table~\ref{tab:varlingam-qc}). A stable model produces responses
to temporary disturbances that eventually decay
\cite{lutkepohl2005new,kilian2017structural}. Same-month checks used
errors from the final causal model, while temporal checks used residuals
from a separately refitted VAR(3). The correlation thresholds were
study-specific choices; their definitions and the bootstrap procedures
are described in Methods. Passing these empirical checks does not verify the full causal identification assumptions or rule out unobserved environmental drivers.

Of the 27,050 successfully fitted pixels, 14,657 (54.18\%) passed all
criteria. All fitted models passed both stability checks, and only
three pixels contained near-constant variables. The joint test of
remaining temporal dependence was the most restrictive criterion:
15,142 pixels (55.98\%) passed. Failures of individual checks overlap,
so each excluded pixel cannot be assigned a single reason for
exclusion.

\begin{table*}[!tp]
\centering
\caption{
    Quality-control results for the per-pixel VAR--LiNGAM models.
    Each diagnostic row reports pixels passing that criterion
    independently; the final row reports pixels passing all criteria.
    Percentages refer to the 27,050 successfully fitted pixels.
}
\label{tab:varlingam-qc}
\small
\begin{tabular}{p{0.36\linewidth}p{0.36\linewidth}rr}
\toprule
Criterion & Pass condition & Pixels & \% \\
\midrule
Fitted-model stability
    & Spectral radius $<1$
    & 27,050 & 100.00 \\
Bootstrap stability
    & At least 95\% of valid fits stable
    & 27,050 & 100.00 \\
Input variability
    & All standard deviations $>10^{-12}$
    & 27,047 & 99.99 \\
Same-month structural-error correlation
    & Maximum absolute correlation $<0.20$
    & 26,275 & 97.13 \\
One-month residual autocorrelation
    & Maximum absolute correlation $<0.30$
    & 27,047 & 99.99 \\
Residual correlations at lags 1--12
    & Maximum absolute correlation $<0.30$
    & 26,942 & 99.60 \\
Joint test of temporal dependence
    & Valid bootstrap test, $p_{\mathrm{boot}}\geq0.05$
    & 15,142 & 55.98 \\
\textbf{Primary analysis pixels}
    & \textbf{All criteria passed}
    & \textbf{14,657} & \textbf{54.18} \\
\bottomrule
\end{tabular}
\end{table*}

\begin{figure}[!tbp]
    \centering
    \includegraphics[width=\linewidth]
        {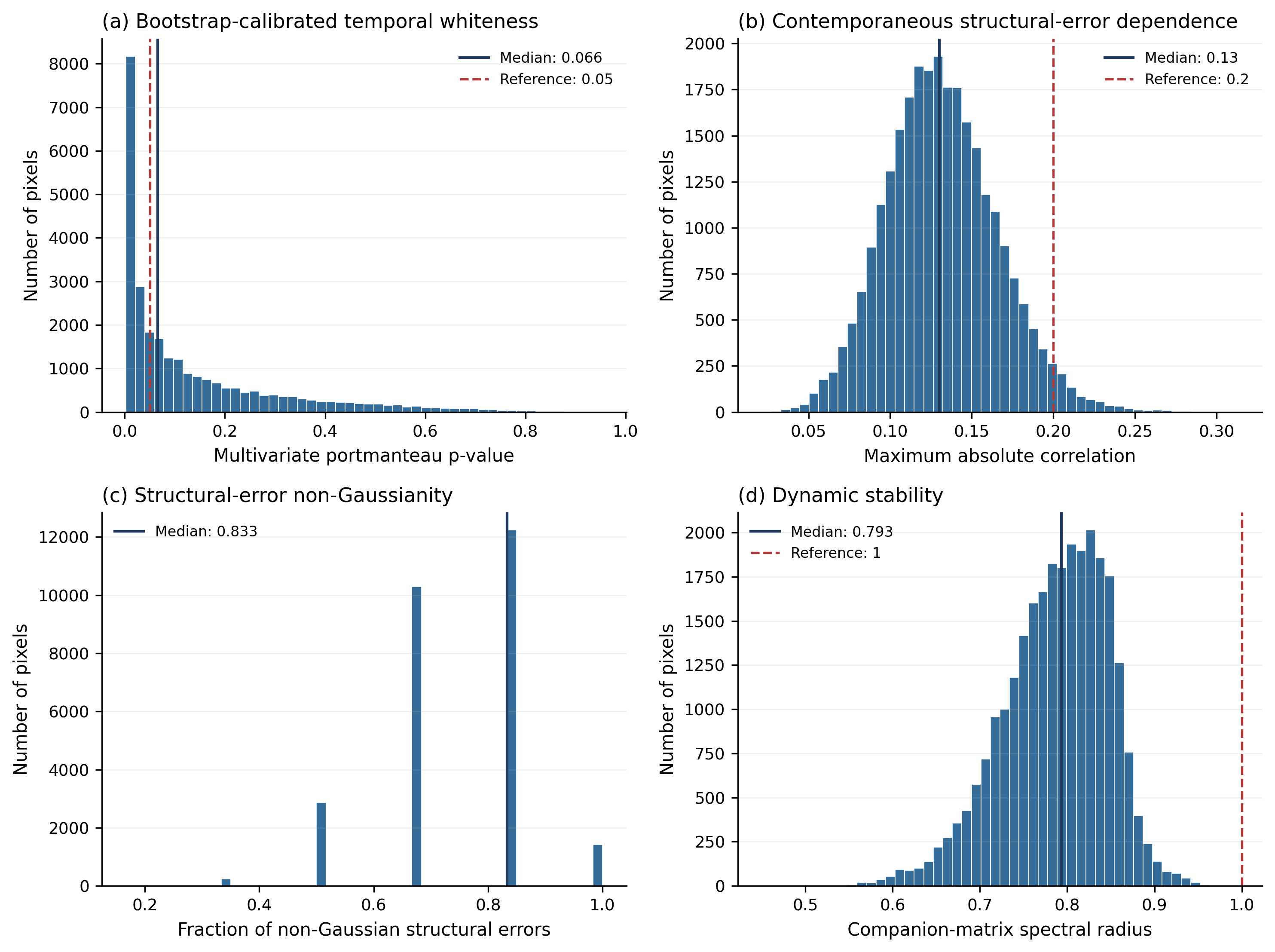}
    \caption{
        Model diagnostics across all successfully fitted pixels,
        before quality-control filtering.
        (a) Bootstrap test of residual temporal dependence through
        lag 12; $p<0.05$ indicates remaining dependence.
        (b) Largest absolute same-month structural-error correlation;
        values below 0.20 passed.
        (c) Fraction of six structural-error series rejecting normality
        at $p<0.05$; this diagnostic was not used for exclusion.
        (d) Dynamic stability; spectral radii below one indicate
        stable models.
    }
    \label{fig:varlingam-diagnostics}
\end{figure}

Figure~\ref{fig:varlingam-diagnostics} summarizes the diagnostics across
all fitted pixels. It also shows the fraction of the six structural-error
series that rejected normality in a Jarque--Bera test. This assessed
the non-Gaussianity assumption of VAR--LiNGAM
\cite{shimizu2006linear}, but was not used to exclude pixels.
Failure to reject normality does not establish that an error series
is Gaussian.

\begin{figure}[!tbp]
    \centering
    \includegraphics[width=\linewidth]
        {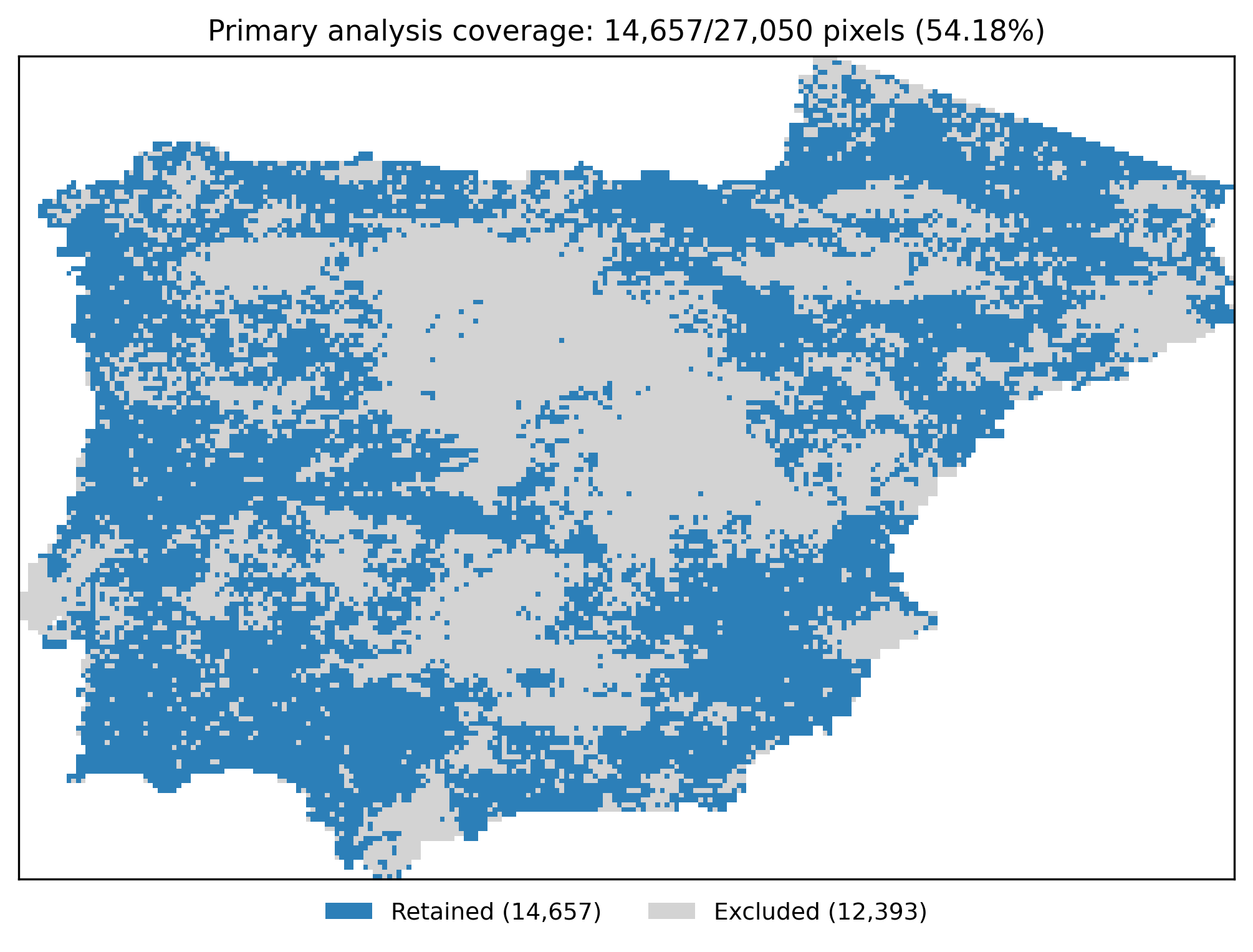}
    \caption{
        Spatial coverage after quality control. The primary analysis
        retained 14,657 of 27,050 fitted pixels (54.18\%).
        Gray pixels failed at least one criterion and were excluded.
    }
    \label{fig:varlingam-qc-coverage}
\end{figure}

The retained pixels were unevenly distributed across the study area
(Figure~\ref{fig:varlingam-qc-coverage}). Gray pixels had fitted models
but failed at least one check. They therefore mark gaps in the primary
analysis, not an absence of vegetation sensitivity. The coverage map
alone cannot explain the geographic pattern of exclusions.

Two plausible explanations for the spatial clustering of QC exclusions
are irrigated agriculture and complex mountain terrain. In irrigated
parts of the Iberian Peninsula, such as the Ebro valley, vegetation
depends on managed water supplies as well as local weather
\cite{kragh2024irrigation}. Irrigation schedules and changes in water
availability could therefore leave temporal patterns that the included
environmental variables do not fully explain, producing different QC
outcomes from nearby rainfed areas. In mountainous areas, elevation
differences and snow accumulation and melt create local conditions
that gridded meteorological inputs may represent incompletely.
Terrain-related limitations, including errors in snow representation,
have been documented for ERA5-Land \cite{munozsabater2021era5land}.
Such mismatches could leave similar temporal error patterns in neighboring
pixels. Together, these mechanisms could explain why coherent areas
fail the temporal-dependence checks while nearby areas pass, although
their contribution would need to be assessed against irrigation and
terrain maps.

\begin{figure}[!tbp]
    \centering
    \includegraphics[width=\linewidth]
        {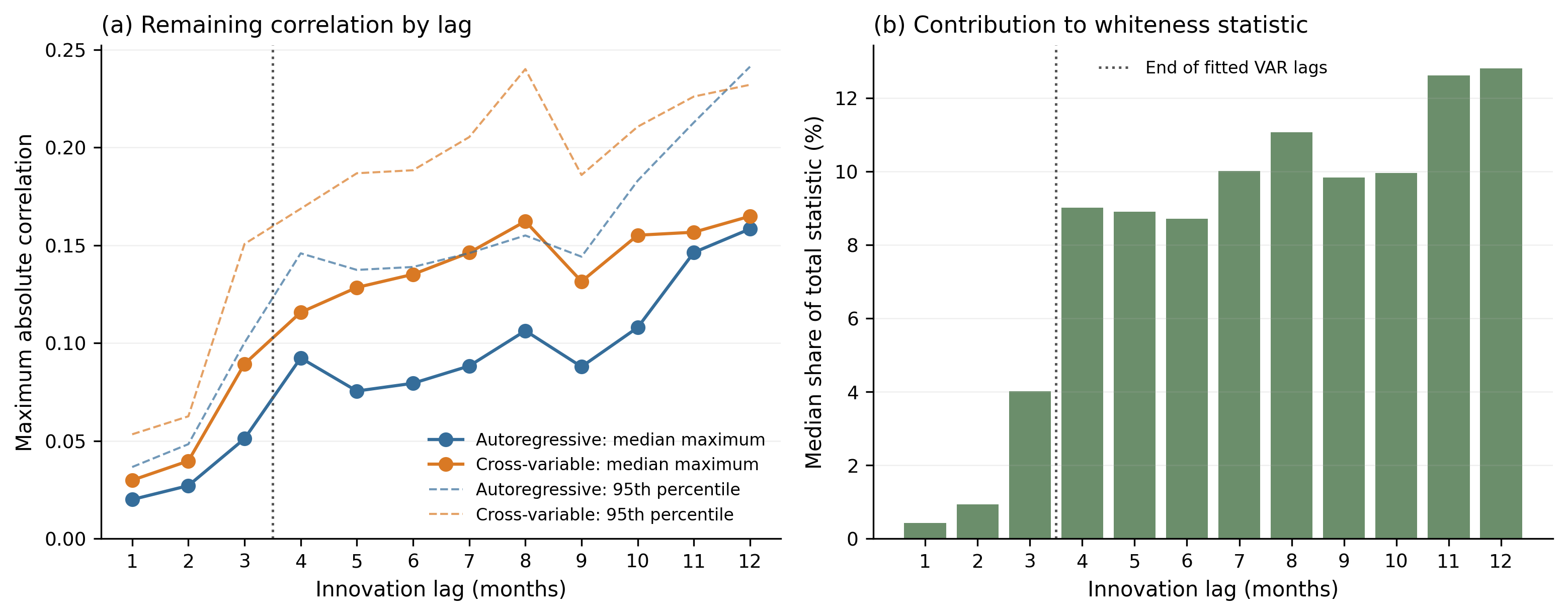}
    \caption{
        Remaining temporal dependence after VAR fitting.
        Left: spatial median and 95th percentile of residual
        correlations within and between variables at each lag.
        Right: relative contribution of each lag to the joint
        temporal-dependence test statistic.
        The vertical marker indicates the three-month lag range
        included in the fitted VAR.
    }
    \label{fig:innovation-lag-profile}
\end{figure}

Figure~\ref{fig:innovation-lag-profile} shows the remaining residual
correlations over 1--12 months. Correlations remained below 0.25 even
at the plotted 95th percentile, and no single lag contributed more
than 12\% of the joint test statistic. Several small correlations can
nevertheless combine to produce a significant test result. Such
patterns may reflect missing drivers, insufficient lag length, or
relationships that vary over time or are nonlinear. These residual
diagnostics do not measure delayed causal effects.

\subsection{Dynamic NDVI effects}

\begin{figure}[!tbp]
  \centering
  \includegraphics[width=\linewidth]
    {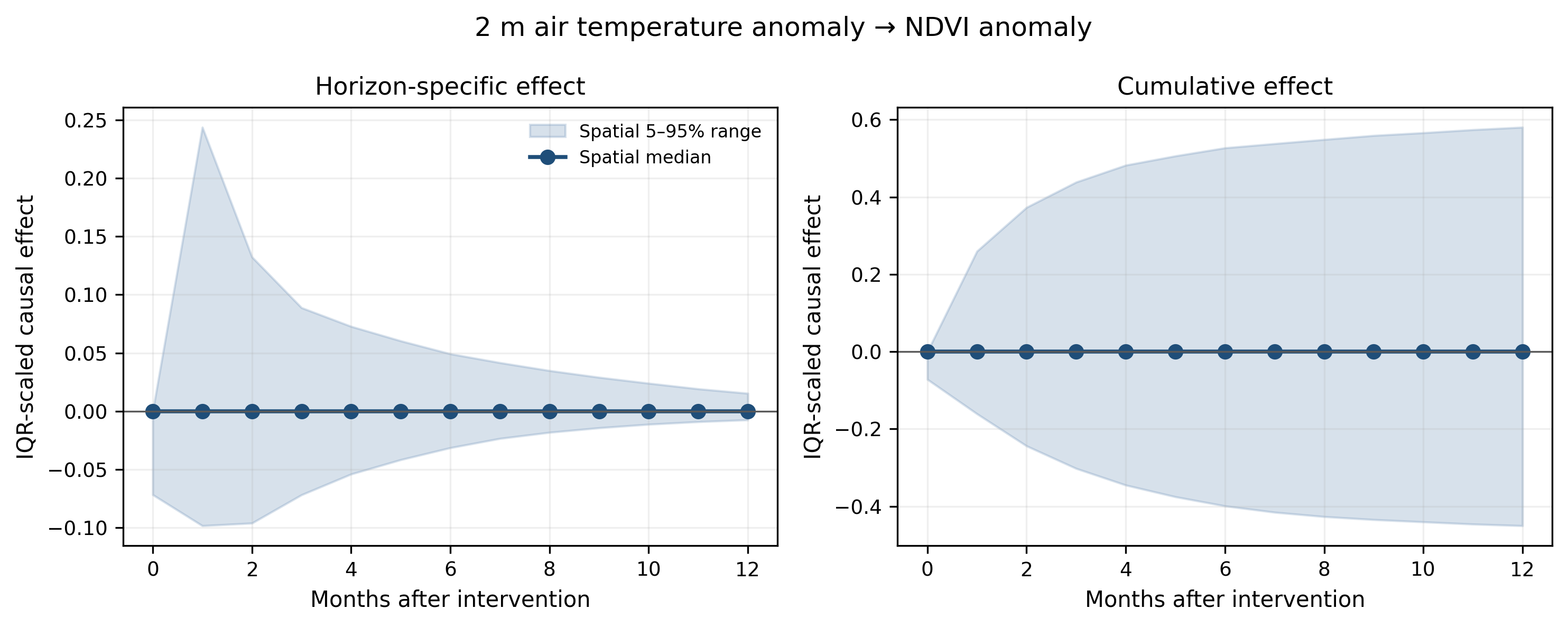}
  \caption{Horizon-specific (left) and cumulative (right) NDVI-anomaly responses to a one-time temperature pulse, scaled by local 10th-to-90th-percentile ranges. The line is the spatial median over 14,657 retained pixels; shading shows their 5th--95th percentile range, not a confidence interval.}
  \label{fig:varlingam-temperature-trajectory}
\end{figure}

\begin{figure}[!tbp]
  \centering
  \includegraphics[width=\linewidth]
    {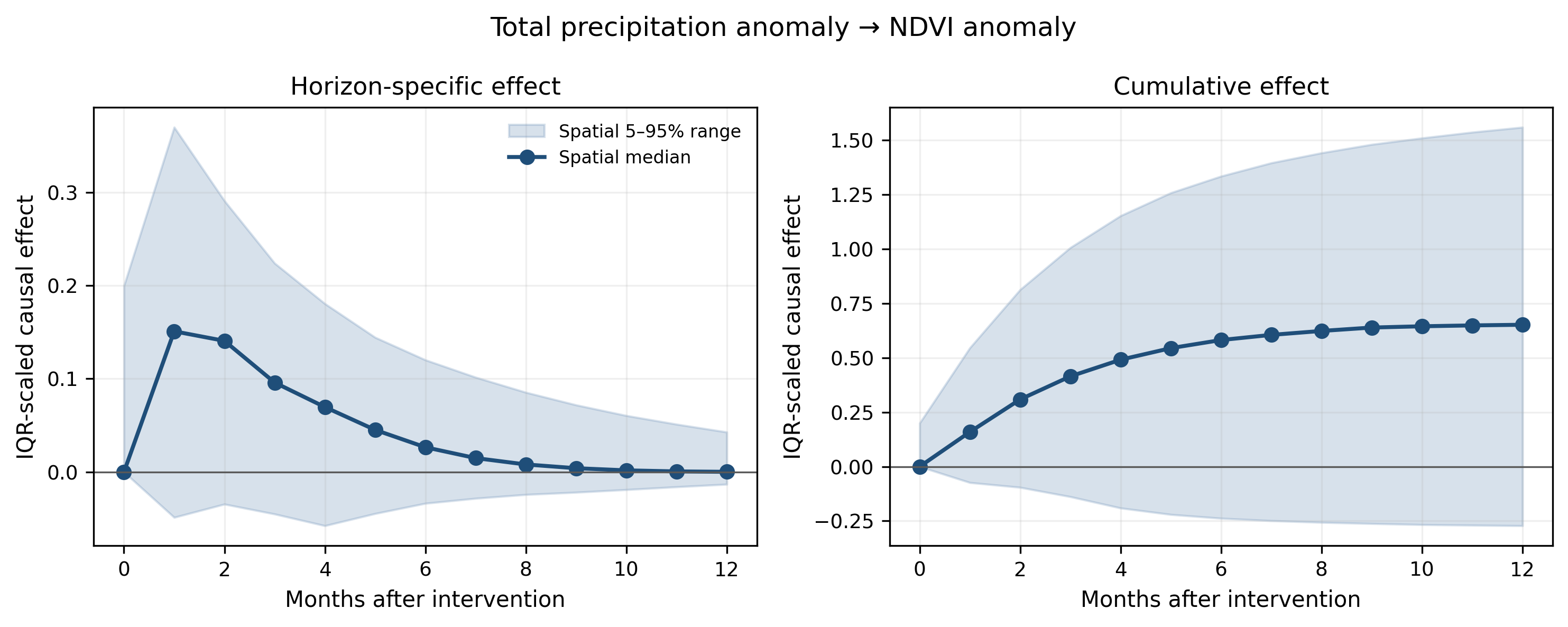}
  \caption{Horizon-specific (left) and cumulative (right) NDVI-anomaly responses to a one-time precipitation pulse, scaled by local 10th-to-90th-percentile ranges. The line is the spatial median over 14,657 retained pixels; shading shows their 5th--95th percentile range, not a confidence interval.}
  \label{fig:varlingam-precipitation-trajectory}
\end{figure}

\begin{figure}[!tbp]
  \centering
  \includegraphics[width=\linewidth]
    {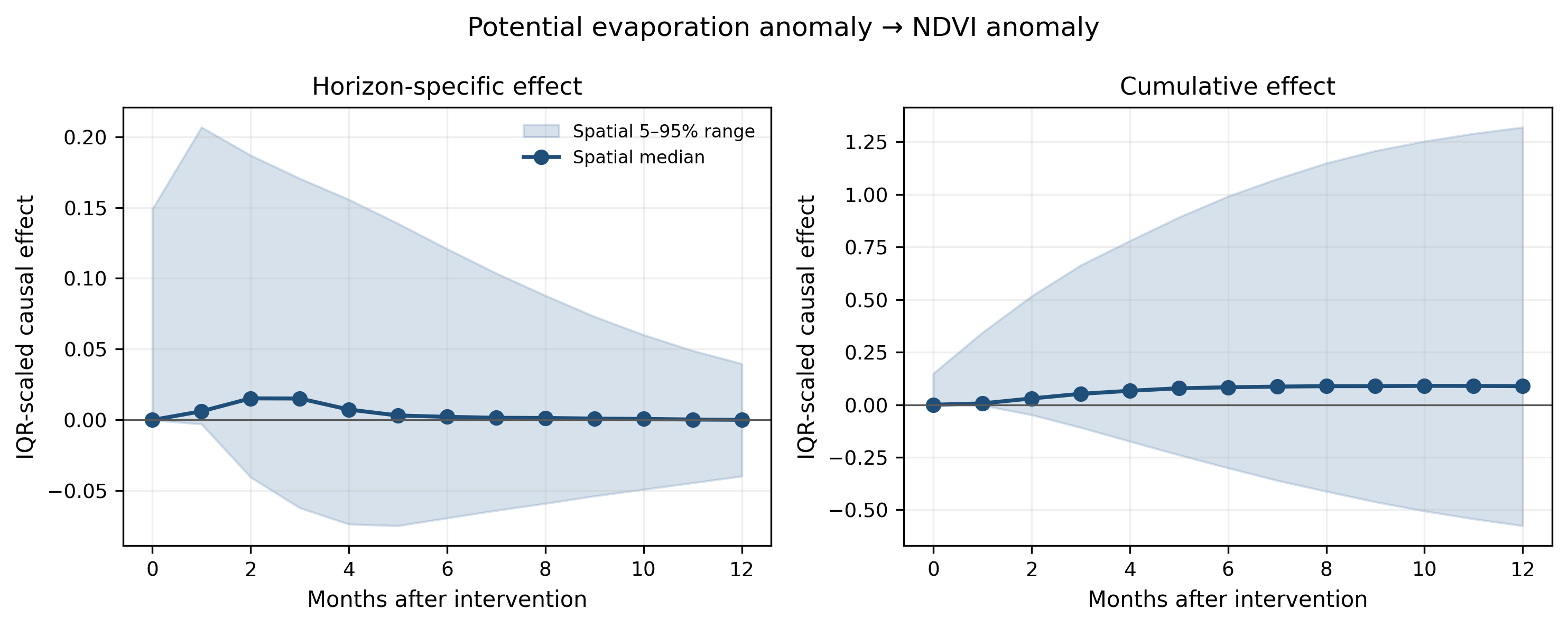}
  \caption{Horizon-specific (left) and cumulative (right) NDVI-anomaly responses to a one-time positive pulse in native ERA5-Land \texttt{pev} (reduced potential evaporation), scaled by local 10th-to-90th-percentile ranges. The line is the spatial median over 14,657 retained pixels; shading shows their 5th--95th percentile range, not a confidence interval.}
  \label{fig:varlingam-evaporation-trajectory}
\end{figure}

\begin{figure}[!tbp]
  \centering
  \includegraphics[width=\linewidth]
    {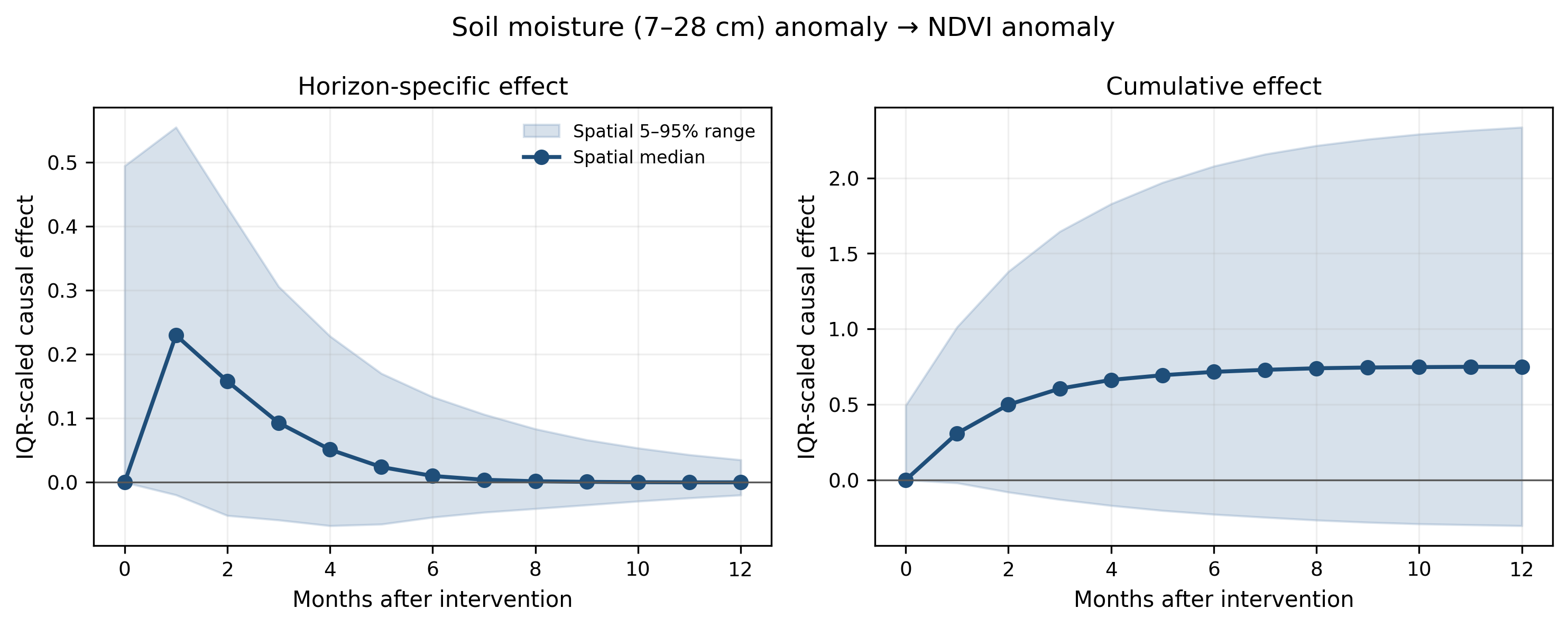}
  \caption{Horizon-specific (left) and cumulative (right) NDVI-anomaly responses to a one-time 7--28\,cm soil moisture pulse, scaled by local 10th-to-90th-percentile ranges. The line is the spatial median over 14,657 retained pixels; shading shows their 5th--95th percentile range, not a confidence interval.}
  \label{fig:varlingam-shallow-soil-moisture-trajectory}
\end{figure}

\begin{figure}[!tbp]
  \centering
  \includegraphics[width=\linewidth]
    {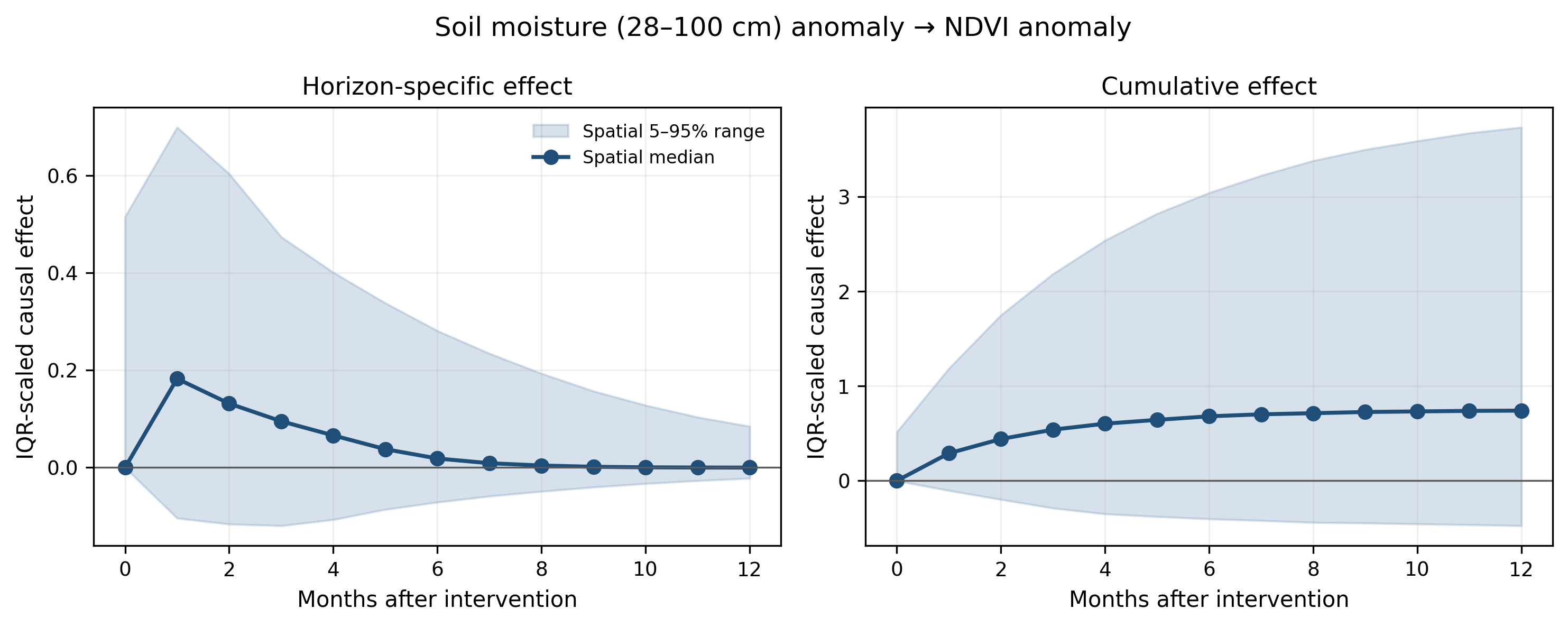}
  \caption{Horizon-specific (left) and cumulative (right) NDVI-anomaly responses to a one-time 28--100\,cm soil moisture pulse, scaled by local 10th-to-90th-percentile ranges. The line is the spatial median over 14,657 retained pixels; shading shows their 5th--95th percentile range, not a confidence interval.}
  \label{fig:varlingam-deep-soil-moisture-trajectory}
\end{figure}

Figures~\ref{fig:varlingam-temperature-trajectory}--\ref{fig:varlingam-deep-soil-moisture-trajectory}
show modeled NDVI-anomaly responses across the 14,657 pixels that passed
primary quality control. Each response follows a one-time positive
pulse to a source equation, with magnitude equal to the local
10th-to-90th-percentile range of the source anomalies. Month 0 represents
the within-month response; subsequent months include indirect pathways
and repeated propagation through the fitted model. Responses can
therefore persist beyond the three explicit VAR lags. The trajectories
describe temporary perturbations rather than permanent changes in
environmental conditions.

The left panels show the response at each horizon, expressed relative
to the local 10th-to-90th-percentile NDVI-anomaly range. A value of 0.20
therefore means that the modeled NDVI response at that horizon equals
20\% of this local range. The right panels sum the responses from
month 0 onwards. Their magnitude combines response strength and
duration: a strong but brief greenness response can produce the same
cumulative value as a smaller response lasting several months.
For example, a cumulative value of 0.60 could represent a response
equal to 20\% of the local NDVI-anomaly range for three months, with
zero response at the remaining horizons. It does not mean that NDVI
is 60\% higher at the final horizon.

At an individual pixel, a cumulative curve that levels off indicates
that the additional monthly response has approached zero. A nonzero
plateau therefore records the accumulated earlier response, rather
than a permanent change in vegetation. Continued accumulation indicates
a persisting response somewhere in the coupled vegetation--environment
system. Conversely, responses that change sign can partly cancel,
producing a small cumulative value despite substantial changes at
individual horizons. Any response continuing beyond month 12 is
outside the mapped accumulation period.

Solid lines represent spatial medians, and shading shows the
5th--95th percentile range across retained pixels. This shading
describes spatial variation (variation across the whole area) rather than uncertainty in the median.
Cumulative responses were calculated separately at each pixel before
spatial summarization, so their median need not equal the sum of the
horizon-specific medians. Likewise, a spatial median near zero can
reflect weak responses, pathways removed during model fitting, or
opposing signs across locations.

\begin{figure*}[!tp]
    \centering
    \includegraphics[width=\textwidth]
        {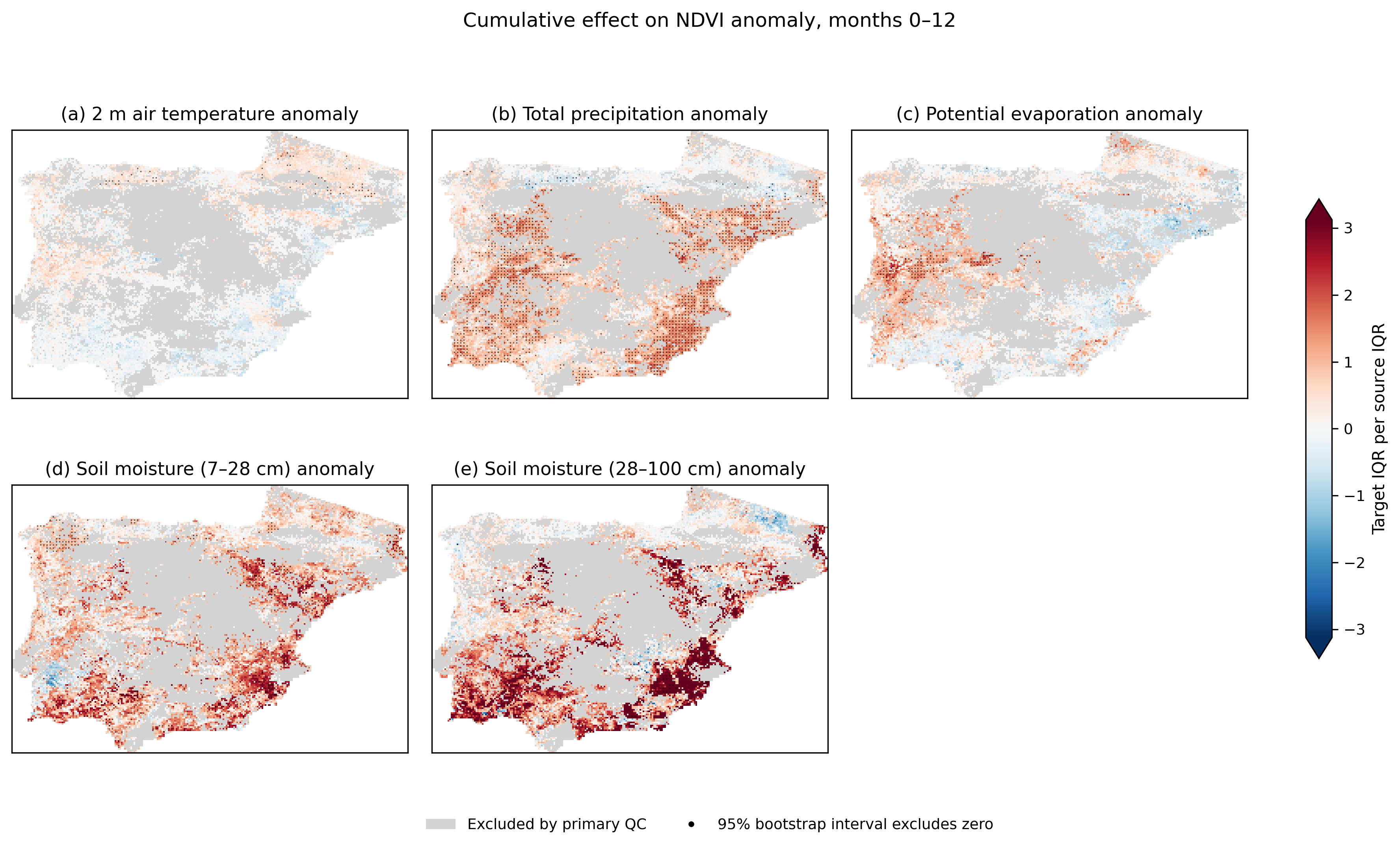}
    \caption{
        Spatial distribution of the cumulative modeled effects of
        precipitation, evaporation, 2-m air temperature, shallow soil
        moisture, and deep soil moisture on NDVI. For each source variable,
        the mapped value is the NDVI response accumulated from month 0 through
        month 12 following a one-time positive pulse in the recorded source
        variable. For native \texttt{pev}, this is a reduction in potential
        evaporation, not an increase. Positive
        values indicate a net increase in NDVI, whereas negative values
        indicate a net decrease. All panels use the same symmetric color
        scale. Gray pixels were fitted but excluded by the primary
        quality-control criteria. Stippling marks pixels where the paired-
        bootstrap 95\% interval for the cumulative effect excludes zero. The
        98th-percentile color limit controls visual clipping only and does not
        alter the estimated effects.
    }
    \label{fig:varlingam-cumulative-effects}
\end{figure*}

Figure~\ref{fig:varlingam-cumulative-effects} maps the cumulative response
from month 0 through month 12, covering 13 monthly response values.
Positive values indicate a net modeled greenness increase over this
window, while negative values indicate a net reduction. For precipitation
and soil moisture, a positive cumulative effect means that the temporary
increase in water input or storage is followed by an accumulated NDVI
increase, including responses transmitted through other variables.
Temperature effects similarly describe the net response to a temporary
warm anomaly, including indirect pathways.

The maps show spatial differences in the magnitude and, for some
sources, the direction of the response. Larger absolute values identify
pixels with a greater accumulated response on the local scale.
However, both the physical pulse size and the NDVI normalisation
vary between pixels. A larger mapped value can therefore reflect
a stronger or more persistent response, a larger source-anomaly range,
or a smaller NDVI-anomaly range. These maps compare responses to
locally scaled perturbations; they do not directly rank sensitivity
to an identical physical perturbation across regions.

The responses describe spectral greenness rather than measured changes
in biomass, carbon uptake, or crop yield. Differences in canopy
structure and the reduced sensitivity of NDVI in dense vegetation
also affect its ecological interpretation \cite{huete2002overview}.
Gray pixels were excluded by primary quality control and should not
be interpreted as having zero response. Stippling identifies pixels
whose paired-bootstrap 95\% interval excludes zero, conditional on the fitted graph model and preprocessing; it is not a spatial multiple-testing-corrected test of a physical intervention.

\subsection{Finite-record recovery under a known structural model}
\label{sec:synthetic-validation-results}

We tested whether residualization and VAR--LiNGAM fitting could recover
known coefficients and vegetation responses from records matching the
240-month observational series. The pruned VAR(3) model from one pixel
passing primary quality control, selected before examining recovery
results, supplied the known contemporaneous and three lagged coefficient
matrices. Structural errors reconstructed from the pixel's observed
residual series were centred and resampled independently across variables
and months. This retained their empirical marginal distributions while
imposing independent structural disturbances. We generated 500 independent
synthetic records, each retaining 240 months after a 200-month burn-in.
The pixel's stored calendar-month effects and linear trend were added to
each simulated series, then re-estimated and removed using the production
residualization procedure. A pruned VAR--LiNGAM model with three lags was
fitted to each resulting residual series.

Recovery was assessed using coefficient errors, edge precision and recall,
sign agreement, and dynamic NDVI responses. For edge comparisons,
unstandardized coefficients with absolute magnitude below 0.01 were treated
as absent. Coefficients were additionally standardized using the original
pixel's residual standard deviations. Dynamic effects were scaled using
the original residual-series 10th--90th-percentile ranges of the source
and target variables, held fixed across simulations. Medians and empirical
5th--95th percentile ranges summarize variation across synthetic records;
these ranges are not confidence intervals for the median.

\begin{figure}[!tbp]
    \centering
    \includegraphics[width=\linewidth]
        {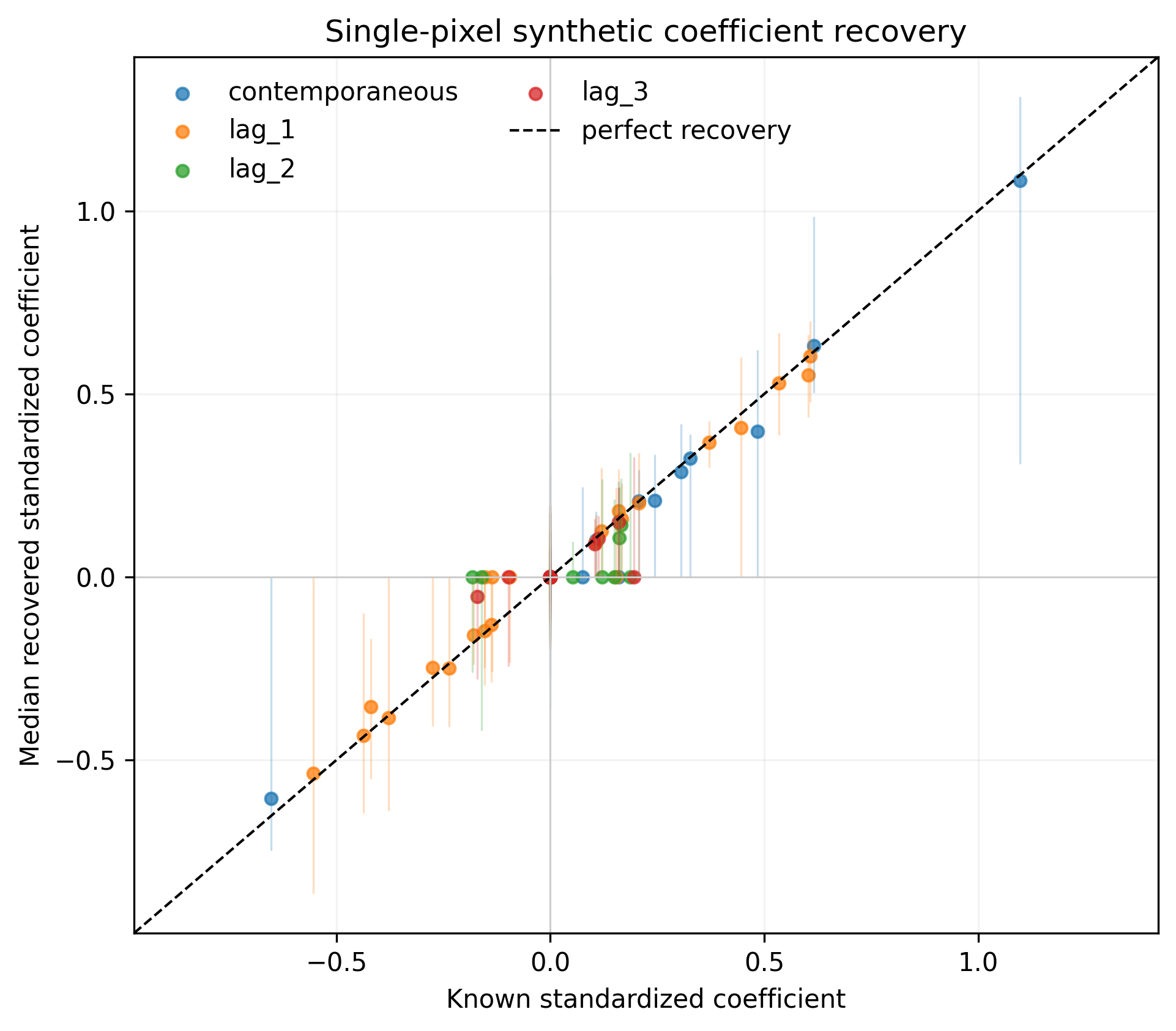}
    \caption{
        Structural-coefficient recovery in the single-pixel synthetic
        experiment. Each point compares a known contemporaneous or
        lagged coefficient with its median estimate across 500
        independently simulated 240-month records. Coefficients are
        standardized using the original pixel's residual standard
        deviations. Vertical lines show the empirical 5th--95th
        percentile range across simulations; the dashed diagonal
        indicates exact agreement.
    }
    \label{fig:synthetic-coefficient-recovery}
\end{figure}

All 500 simulations produced successful, dynamically stable fits
(Table~\ref{tab:synthetic-recovery}). Median standardized coefficient
estimates broadly followed the known values
(Figure~\ref{fig:synthetic-coefficient-recovery}). The median correlation
between known and recovered unstandardized coefficients was 0.957,
although its 5th--95th percentile range of 0.021--0.995 showed substantial
variation between records. Thus, successful fitting and dynamic stability
did not guarantee accurate coefficient recovery. Median standardized RMSE
was 0.080 across all evaluated coefficients and 0.118 among true edges,
showing that including absent or very small coefficients lowered the
overall error.

Median edge precision, recall, and F1 were 0.857, 0.690, and 0.763,
respectively. Most identified edges were therefore present in the
generating model, while approximately 31\% of the threshold-defined true
edges were missed in a typical realization. Among correctly recovered
edges, median sign agreement was 1.000, with a 5th--95th percentile range
of 0.965--1.000.

\begin{figure*}[!tp]
    \centering
    \includegraphics[width=\textwidth]
        {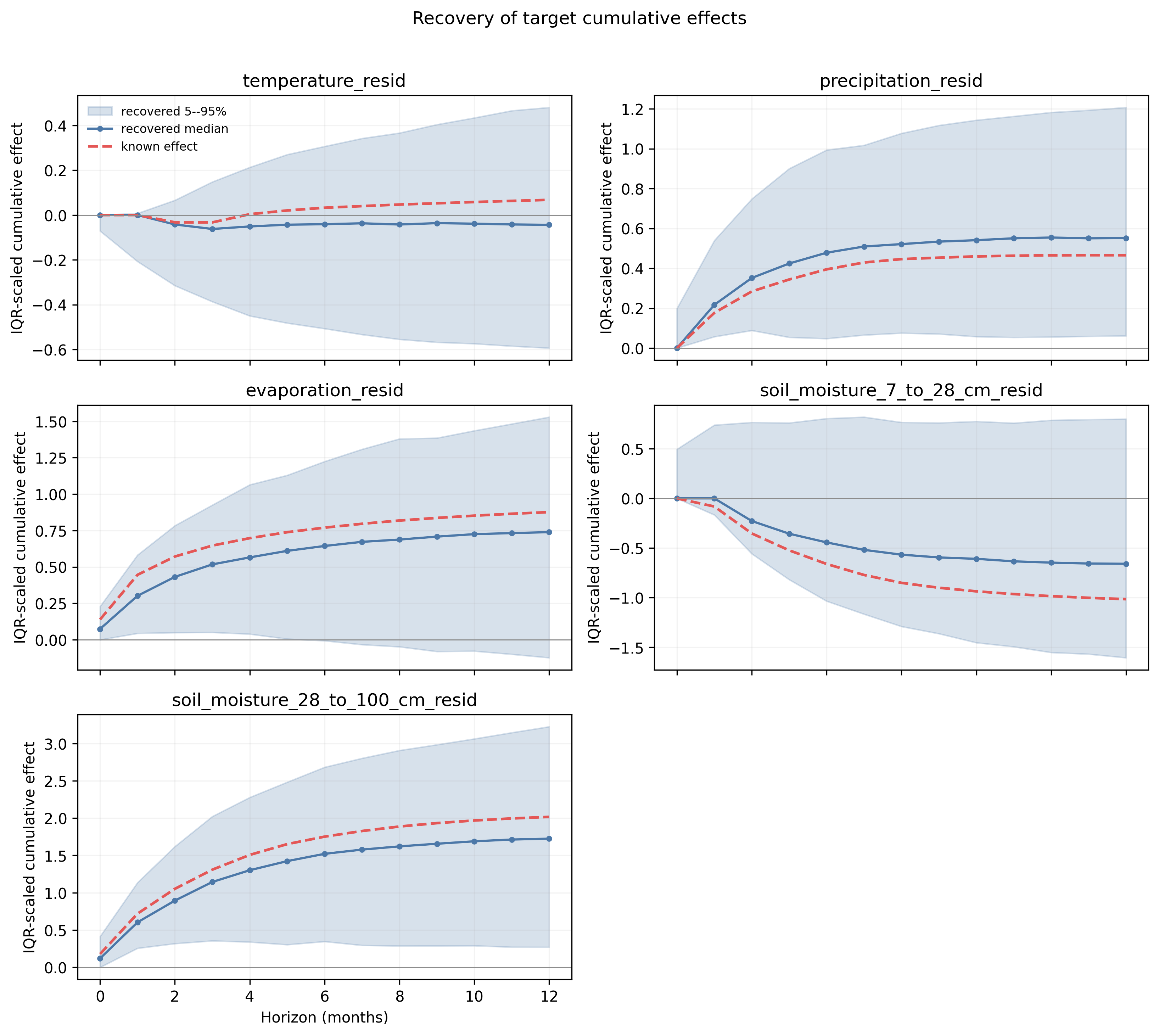}
    \caption{
        Recovery of cumulative NDVI responses to one-time pulses in
        the five environmental drivers. Dashed lines show responses
        implied by the generating model, solid lines show median
        recovered responses, and shading shows the empirical
        5th--95th percentile range across 500 independently simulated
        240-month records. Each trajectory sums responses from
        month 0 through the indicated horizon. Pulse amplitudes equal
        the source variable's local residual-series 10th--90th-percentile
        range, and NDVI responses are expressed relative to the
        corresponding NDVI range. These scales are taken from the
        original pixel and held fixed across simulations. Shading
        describes variation between simulated records, conditional
        on the generating model.
    }
    \label{fig:synthetic-cumulative-effect-recovery}
\end{figure*}

Median recovered cumulative NDVI trajectories broadly followed the known
trajectories (Figure~\ref{fig:synthetic-cumulative-effect-recovery}), but
the recovery ranges were wide for several drivers and horizons. At
horizon 12, the median cumulative-effect RMSE was 0.553 in locally scaled
units, with a 5th--95th percentile range of 0.249--1.130. The median
residualization-baseline RMSE was 0.236 original residual standard
deviations. Agreement of the median trajectories therefore coexisted
with substantial uncertainty for an individual 240-month record.

\begin{table}[!tbp]
    \centering
    \small
    \caption{
        Recovery statistics for the single-pixel synthetic experiment.
        Values are medians with empirical 5th--95th percentile ranges
        in parentheses, except for fit and stability rates, which
        describe the proportion of the 500 simulations. True edges
        satisfy $|B|\geq0.01$ on the unstandardized coefficient scale.
        Sign agreement is evaluated only among correctly recovered
        edges.
    }
    \label{tab:synthetic-recovery}
    \begin{tabular}{@{}p{0.56\columnwidth}p{0.38\columnwidth}@{}}
        \toprule
        \textbf{Recovery metric} & \textbf{Estimate} \\
        \midrule
        Successful model fits
            & 100.0\% \\
        Dynamically stable recovered models
            & 100.0\% \\
        Coefficient correlation (unstandardized)
            & 0.957 (0.021--0.995) \\
        Standardized coefficient RMSE, all coefficients
            & 0.080 (0.060--0.154) \\
        Standardized coefficient RMSE, true edges
            & 0.118 (0.087--0.226) \\
        Edge precision
            & 0.857 (0.742--0.945) \\
        Edge recall
            & 0.690 (0.500--0.833) \\
        Edge F1 score
            & 0.763 (0.626--0.864) \\
        Sign agreement among correctly recovered edges
            & 1.000 (0.965--1.000) \\
        Cumulative-effect RMSE through month 12
            & 0.553 (0.249--1.130) \\
        Baseline RMSE (residual standard deviations)
            & 0.236 (0.181--0.331) \\
        \bottomrule
    \end{tabular}
\end{table}

In this single-pixel, model-based benchmark, 240-month records recovered some broad response patterns but showed substantial realization-to-realization variation. Graph fitting and dynamic stability were therefore insufficient indicators of accurate causal recovery by themselves. This motivates the use of bootstrap uncertainty and local diagnostics in the observational analysis. Because the generating model was derived from one fitted pixel, the experiment evaluates finite-record behavior under that model rather than establishing the correctness of the observed Iberian causal graphs.

\subsection{External climatic evaluation of spatial vegetation responses}

\begin{figure*}[!tp]
    \centering
    \includegraphics[width=\textwidth]
        {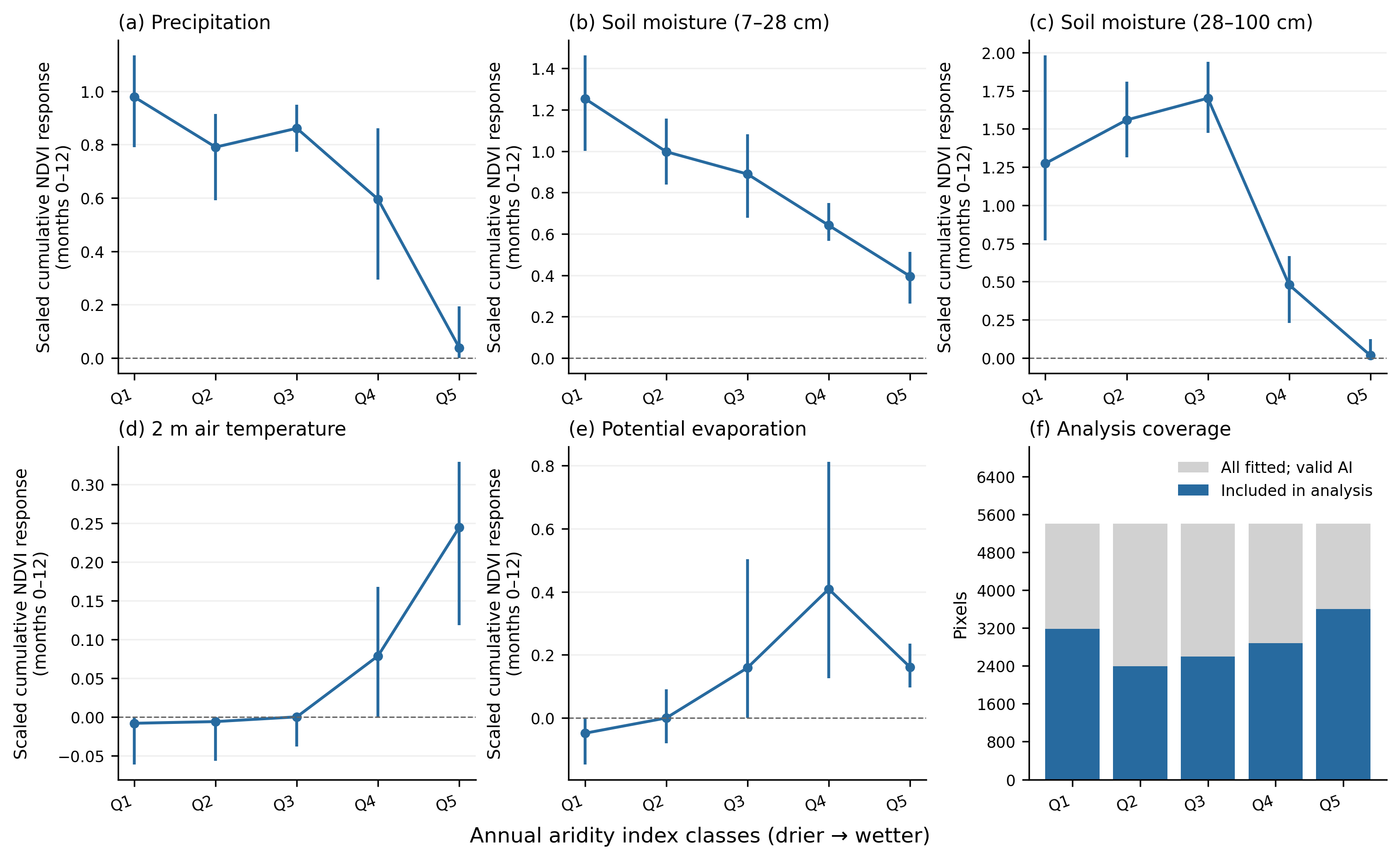}
    \caption{
        Vegetation responses along the external aridity gradient,
        from the driest quintile (Q1) to the wettest (Q5).
        Panels (a--e) show median NDVI-anomaly responses accumulated
        over months 0--12 following a one-time positive source pulse.
        Pulse amplitudes equal the local source residual-series
        10th--90th-percentile range, and NDVI responses are expressed
        relative to the corresponding NDVI range. Effects retain
        the input-variable sign conventions. Vertical lines show
        pointwise 95\% spatial-block bootstrap confidence intervals,
        conditional on the fitted effects. Panel (f) shows fitted
        pixels with valid AI (gray) and those retained after quality
        control and the complete-effect requirement (blue).
        Quintile boundaries were defined before quality control.
    }
    \label{fig:aridity-signed-effects}
\end{figure*}

\begin{table*}[!tp]
    \centering
    \small
    \caption{
        Associations between aridity and fitted vegetation responses
        across 100\,km geographic blocks. Spearman correlations compare
        block-median AI with block-median response; higher AI indicates
        wetter conditions. Source-specific responses are signed,
        locally scaled cumulative NDVI effects over months 0--12.
        Permutation $p$-values use 4,999 permutations.
        Benjamini--Hochberg adjustment applies to the five
        source-specific tests; the dash denotes that it was not
        applied to the separate primary water-response-share test.
    }
    \label{tab:aridity-trends}
    \begin{tabular}{lrrrr}
        \toprule
        Response & Blocks & Spearman $\rho$
            & $p_{\mathrm{perm}}$ & $p_{\mathrm{BH}}$ \\
        \midrule
        Water-response share
            & 85 & -0.705 & 0.0002 & -- \\
        Precipitation
            & 85 & -0.620 & 0.0002 & 0.0003 \\
        Soil moisture (7--28\,cm)
            & 85 & -0.632 & 0.0002 & 0.0003 \\
        Soil moisture (28--100\,cm)
            & 85 & -0.700 & 0.0002 & 0.0003 \\
        2\,m air temperature
            & 85 & 0.691 & 0.0002 & 0.0003 \\
        Potential evaporation
            & 85 & 0.328 & 0.0028 & 0.0028 \\
        \bottomrule
    \end{tabular}
\end{table*}

We evaluated whether the geographical variation in fitted vegetation responses was consistent with an independent measure of long-term climatic water availability, using the independently produced Global-AI v3 aridity
climatology for 1970--2000 \cite{zomer2022aridity}. Its annual aridity index
($AI=P/ET_0$) increases from dry to humid conditions. We averaged AI within
each analysis-pixel footprint, requiring at least 50\% valid coverage,
and defined quintiles across all fitted pixels with valid AI. Response
summaries included pixels passing primary quality control with complete
estimates for all five sources. We compared their signed, locally scaled
cumulative NDVI responses over months 0--12.

To summarize the relative magnitude of water-related and atmospheric
responses, we calculated
$S_i=(W_i/3)/(W_i/3+A_i/2)$, where $W_i$ sums the absolute cumulative
responses to precipitation and both soil-moisture variables, and $A_i$
sums those for temperature and potential evaporation. Dividing by the
number of sources gives equal weight to the mean response magnitude
in each group. The index was omitted where both groups had zero response.
Quintile medians were accompanied by 95\% confidence intervals from
999 spatial-block bootstrap samples within each quintile. We assessed
associations with AI using Spearman correlations between median AI and
median responses within 100\,km geographic blocks, with 4,999 permutations.
The five source-specific tests received Benjamini--Hochberg adjustment.
Published aridity classes and block sizes of 50 and 150\,km provided
sensitivity analyses.

Median cumulative responses to precipitation and both soil-moisture
variables decreased from the driest to the wettest quintile
(Figure~\ref{fig:aridity-signed-effects}). Temperature responses increased,
while the response to a positive native-\texttt{pev} pulse (reduced potential evaporation) showed a less consistent increase, departing from the general pattern in the wettest quintile. The block-level
analysis found all five associations statistically significant after adjustment for multiple
testing ($p_{\mathrm{BH}}\leq0.0028$;
Table~\ref{tab:aridity-trends}). These correlations describe changes in
signed responses; their magnitudes indicate the consistency of the
climatic gradient, rather than the size of the NDVI effect.

Water-response share decreased strongly toward wetter conditions across the 85 geographic blocks (\(\rho=-0.705\), \(p_{\mathrm{perm}}=0.0002\)), indicating that water-related responses were relatively stronger in drier environments. The corresponding source-specific trends were consistent with this pattern: precipitation and both soil-moisture responses declined with increasing aridity index, whereas temperature response increased. Agreement with the external climatology indicates a physically interpretable climatic pattern, but does not establish whether it is caused by the mechanisms represented in the fitted graphs. Identification at individual pixels remains conditional on the VAR--LiNGAM assumptions and the measured variables.

\FloatBarrier
\section{Discussion}

The Iberian case study shows marked geographic variation in locally scaled, model-implied NDVI responses to hydroclimatic disturbances. In the retained sample, the relative contribution of water-related responses declines along the external aridity gradient, while the signed temperature response tends to increase toward wetter conditions. This agreement with a long-term climatological gradient is consistent with plausible vegetation--water relationships, but it is not independent verification of individual causal edges. Nor does this analysis demonstrate that a single region-wide model would perform worse: no regional-model comparison was conducted.

Model adequacy also varies spatially. Although the fitted models were dynamically stable, only 14,657 of 27,050 pixels (54.18\%) passed the full primary screening criteria. The bootstrap test of residual temporal dependence was the most restrictive criterion. In particular, the temporal test assesses a refitted reduced-form VAR rather than every aspect of the final pruned structural model. Rejection signals that residual temporal structure remains under the tested specification, not that a specific edge or fitted NDVI response has been proven false. The resulting QC mask should therefore be read as an empirical scope restriction rather than a validated map of causal identifiability.

The structural assumptions remain consequential even for pixels passing these tests. Linear, time-invariant relations and acyclic same-month interactions may only approximate ecohydrological dynamics; the included variables do not capture irrigation operations, land management, snow dynamics, or all relevant human and environmental drivers. ERA5-Land soil moisture and potential evaporation are model-based products rather than independent in-situ measurements, and their errors or shared modeling assumptions could influence the recovered relations. Residualization and pixel-wise fitting do not remove hidden time-varying common causes, and the analysis does not represent spillovers between neighboring pixels. The bootstrap intervals further condition on the preprocessing, measured products, selected lag order, and assumed model class.

The synthetic experiment characterizes estimation under one fitted generating system. Its broad median recovery and wide record-level ranges justify reporting finite-sample uncertainty, but they cannot establish recovery across all Iberian landscapes or prove that any observational graph is correct. The external aridity assessment likewise provides a useful consistency check, not a test of intervention responses. Independent field records, product cross-checks, or intervention-based designs would be required for stronger attribution; these lie outside the present study. The main contribution is therefore to place spatially varying structural response estimates alongside explicit uncertainty and model diagnostics.

\section{Conclusion}

Across the Iberian Peninsula, the estimated effects of climate and soil moisture on vegetation differed considerably in both magnitude and duration. Of the 27,050 local VAR–LiNGAM models fitted to 20 years of monthly observations, 14,657 met the diagnostic criteria for further analysis. These models revealed a clear geographical pattern: water-related variables had a greater relative influence on vegetation in drier areas, while temperature effects became more positive toward wetter conditions. The agreement with an independent aridity climatology suggests that the models capture meaningful differences in vegetation sensitivity, although it does not establish the validity of individual causal relationships.

The results also highlight the limitations of estimating causal relationships independently from relatively short environmental time series. Our synthetic experiment showed substantial variation in recovered effects between realizations, while almost half of the observational models failed at least one diagnostic check. We plan to address this problem through partial pooling approaches that allow information to be shared between locations without imposing a single causal model across the study area. Rather than estimating each pixel in isolation, such methods could draw on similarities between locations to improve the stability of causal discovery and effect estimation, while retaining differences supported by the local observations. An important question will be how much information can be shared before genuine geographical variation begins to be obscured. Partial pooling cannot resolve violations of the underlying causal assumptions, but it offers a promising way to overcome some of the statistical limitations of short EO time series. Developing and evaluating these approaches will be the next step in extending Confoundry to more reliable spatial causal analysis.

\IEEEtriggeratref{29}
\renewcommand{\IEEEbibitemsep}{-1.25pt}
\bibliographystyle{IEEEtran}
\bibliography{references}

\end{document}